\documentclass[final]{siamart0216}

\usepackage{graphicx}  
\usepackage{epstopdf}
\usepackage{algorithmic}
\ifpdf
  \DeclareGraphicsExtensions{.eps,.pdf,.png,.jpg}
\else
  \DeclareGraphicsExtensions{.eps}
\fi

\usepackage{caption}
\usepackage{subcaption}
\usepackage{enumerate}
\usepackage{enumitem}

\newcommand{\rmnum}[1]{\romannumeral #1}

\newcount\Comments  
\usepackage{color}
\definecolor{darkgreen}{rgb}{0,0.5,0}
\definecolor{purple}{rgb}{1,0,1}
\newcommand{\kibitz}[2]{\ifnum\Comments=0\textcolor{#1}{#2}\fi}

\newcommand{\edit}[1] {\kibitz{black}     {#1}}

\newcommand{\TheTitle}{A convex formulation of traffic dynamics on transportation networks} 
\newcommand{\TheAuthors}{Yanning Li, Christian Claudel, Benedetto Piccoli, Daniel B. Work}

\headers{\TheTitle}{\TheAuthors}

\title{{\TheTitle}}

\author{ \small
  Yanning Li\thanks{PhD Student in the Department of Civil and Environmental Engineering, University of Illinois at Urbana Champaign,  Urbana, IL 61801, USA.
    (\email{yli171@illinois.edu}).}
  \and
  Christian G. Claudel\thanks{Assistant Professor in the Department of Civil, Architectural and Environmental Engineering, University of Texas at Austin, Austin, TX 78712, USA. 
  (\email{christian.claudel@utexas.edu}).}
  \and
 Benedetto Piccoli \thanks{Professor in the Department of Mathematical Science, Rutgers University-- Camden, Camden, NJ 08102, USA.
  (\email{piccoli@camden.rutgers.edu}).}
  \and
  Daniel B. Work\thanks{Corresponding author. Assistant Professor in the Department of Civil and Environmental Engineering, and Coordinated Science Laboratory, University of Illinois at Urbana Champaign,  Urbana, IL 61801, USA.
  (\email{dbwork@illinois.edu}).}
}

\usepackage{amsopn}

\ifpdf
\hypersetup{
  pdftitle={\TheTitle},
  pdfauthor={\TheAuthors}
}
\fi

\begin{document}

\maketitle

\begin{abstract}                        
This article proposes a numerical scheme for computing the evolution of vehicular traffic on a road network over a finite time horizon. The traffic dynamics on each link is modeled by the \emph{Hamilton-Jacobi} (HJ) \emph{partial differential equation} (PDE), which is an equivalent form of the \textit{Lighthill-Whitham-Richards} PDE. The main contribution of this article is the construction of a single convex optimization program which computes the traffic flow at a junction over a finite time horizon and decouples the PDEs on connecting links. Compared to discretization schemes which require the computation of all traffic states on a time-space grid, the proposed convex optimization approach computes the boundary flows at the junction using only the initial condition on links and the boundary conditions of the network. The computed boundary flows at the junction specify the boundary condition for the HJ PDE on connecting links, which then can be separately solved using an existing semi-explicit scheme for single link HJ PDE. As demonstrated in a numerical example of ramp metering control, the proposed convex optimization approach also provides a natural framework for optimal traffic control applications.
\end{abstract}





\section{Introduction}
\label{s:intro}
\edit{Efficient numerical schemes for solving traffic dynamics modeled by a conservation law on a transportation network are critical and the basis for traffic applications, such as real-time traffic estimation and optimal traffic control. This article proposes a numerical scheme which solves the traffic dynamics in a convex program without the discretization of the time-space domain and can be used as the framework for optimal control on transportation networks.}

The road network is represented by a directed graph $\mathcal{G}(\mathcal{L}, \mathcal{V})$ consisting of links $l \in \mathcal{L}$ and vertices $v \in \mathcal{V}$. Each link $l$ represents a road segment with spatial coordinates  $x\in [a_{l},\, b_{l}]$ and homogeneous physical parameters, such as the free flow speed and the capacity. Each vertex $v$ represents a junction in the transportation network, consisting of at least one incoming link and one outgoing link. The link endpoints that are not connected to a junction are referred to as the \emph{network boundaries} while endpoints at junctions are called \emph{internal boundaries}.

Describing the dynamics of traffic on a network consists of two modeling components, namely a model for the traffic evolution on each link, and a model of traffic flow through each junction. The standard first order model for traffic flow on a single link indexed by $l$ is the \emph{Lighthill-Whitham-Richards} (LWR) \emph{partial differential equation} (PDE)~\cite{TrafficNetwork:lighthill1955kinematic, TrafficNetwork:richards1956shock}:
\begin{equation}\label{e:lwr}
{\frac{\partial \rho_{l} (t,x)}{\partial t}} + {\frac{\partial{ \psi_l\left(\rho_{l}(t,x)\right) }}{\partial x}}=0,
\end{equation}
which describes the evolution of the traffic density $\rho_{l}(t,x)$ in the time and space domain $[0,t_{max}]\times [a_l,b_l]$. In \eqref{e:lwr},  $\psi_l(\cdot)$ is the flux function which describes the empirical relationship between the density and flow on each link. The initial and boundary conditions are respectively defined as $\rho_{l}(0, x) = \rho_{l,0}(x), x \in [a_{l}, b_{l}]$ and $\rho_{l}(t,a_l) = \rho_{a_l}(t),\, \rho_{l}(t,b_l) = \rho_{b_l}(t), t \in [0, t_{max}]$. 

There is an increasing interest in an equivalent representation of  the LWR PDE~\eqref{e:lwr} which is known as the \emph{Hamilton-Jacobi} (HJ) PDE~\cite{aubin2008dirichlet, claudel2010lax1, claudel2010lax2, Daganzo05, daganzo2005variational, daganzo2006variational}. By integrating~\eqref{e:lwr} in space $x$, the following HJ PDE can be obtained:
\begin{equation}
\label{e:hjpde}
\frac{\partial {\bf M}_{l}(t,x)}{\partial t}
    -\psi_l\left(-\frac{\partial {\bf M}_{l}(t,x)} {\partial x}\right)
\; = \;  0.
\end{equation}
In formulation~\eqref{e:hjpde}, the traffic state is described by a real scalar function $\mathbf{M}_{l}(t,x)$, known as the Moskowitz function~\cite{TrafficNetwork:moskowitz1965discussion, newell1993simplified_part1, newell1993simplified_part2, newell1993simplified_part3}. Intuitively, $\mathbf{M}_{l}(t,x)$ is a continuous analog of a sequentially indexed vehicle ID (or cumulative vehicle count) and is related to the density by \edit{$\rho_{l}(t,x) = -{\partial \mathbf{M}_{l}(t,x)}/{\partial x}$}. All vehicles (including vehicles on the road at $t=0$) are labeled incrementally in the order they enter the link, where negative labels are assigned to vehicles initially on the link. As a convention, the continuous analog of the vehicle ID at $(t,x)=(0,a_{l})$ is set as $\mathbf{M}_{l}(0,a_{l}) = 0$. The initial and boundary conditions are respectively $\mathbf{M}_l(0,x) = \mathbf{M}_{l,0}(x),\; \mathbf{M}_l(t,a_l)= \mathbf{M}_{a_l}(t),\; \mathbf{M}_l(t,b_l)= \mathbf{M}_{b_l}(t)$, where the boundary Moskowitz data is related to the density data by $\mathbf{M}_{l,0}(x) =  -\int_{\chi=a_l}^{x} $ $\rho_{l, 0}(\chi) {\rm d}\chi$, $\mathbf{M}_{a_l}(t) = \int_{\tau=0}^{t} \psi_l\left(\rho_{a_{l}}(\tau)\right) {\rm d}\tau$, and $\mathbf{M}_{b_l}(t) = \mathbf{M}_{l,0}(b_l) + \int_{\tau=0}^{t} \psi_l\left(\rho_{b_{l}}(t)\right){\rm d}\tau$. The HJ PDE formulation~\eqref{e:hjpde} enables the use of variational theory~\cite{Daganzo05} to compute the vehicle ID $\mathbf{M}_l$ at any point $(t,x)$ by minimizing a functional given a concave Hamiltonian $\psi_l(\cdot)$ (i.e., a concave flux function).

The classic numerical schemes for computing the traffic evolution on a single link are based on discretization of the governing PDE. These schemes include the Godunov Scheme~\cite{godunov1959scheme} and the discrete velocities kinetic scheme~\cite{bretti2006numerical} for the LWR PDE~\eqref{e:lwr}, or the dynamic programming approach~\cite{Daganzo05} for the HJ PDE~\eqref{e:hjpde}. Recently, Claudel and Bayen~\cite{claudel2010lax1, mazare2011analytical} proposed a semi-explicit HJ PDE solver based on the \emph{Lax-Hopf} formula for a single link, which has been demonstrated to be more efficient than discretization-based schemes~\cite{claudel2010lax2}. Built upon the semi-explicit HJ PDE solver on single link, this article develops a numerical scheme for extending the HJ PDE solver to a network.

To extend the link traffic flow model to networks, a junction model is required to describe how the traffic sent from links $s\in\mathcal{S}_{v}\subset{\mathcal{L}}$ entering the junction $v$ is received by links $r\in\mathcal{R}_{v}\subset{\mathcal{L}}$ exiting the junction at any point in time. It is well known that conservation of vehicles across the junction, i.e., $\sum_{s\in \mathcal{S}_v} q_s(t, b_s) = \sum_{r \in \mathcal{R}_v} q_r(t, a_r) $, is insufficient to uniquely define the flows at the junction. To address this issue, a variety of junction models~\cite{daganzo1995net, piccoli2006traffic, han2015continuous, herty2003modeling, holden1995mathematical, jin2010continuous, jin2012riemann, jin2003distribution, lebacque2005intersection} have been proposed to define a unique internal boundary flow solution using additional rules governing the distribution or priority of the flows. \edit{Compared to the merge junction, for which relatively few models have been proposed, the diverge junctions have led to a number of modeling efforts. The diverge models can be classified as \textit{First-In-First-Out} (FIFO) and non-FIFO models. The FIFO model is directly applicable to single-lane roadways while the non-FIFO model can be applied to multi-lane scenarios.} In this article, we adopt a merge junction model~\cite{piccoli2006traffic,han2015continuous} and propose a new \edit{(FIFO)} diverge junction model which allows rerouting behaviors while maintaining consistency (or equivalently invariance~\cite{lebacque2005first}). Moreover, both junction models allow the junction flows to be computed pointwise in time as a convex optimization program, which is an important feature used in the proposed numerical scheme. It should be noted that the computed boundary flow values can be posed as the strong boundary condition to the PDEs which is commonly discussed using the concept of Riemann solver~\cite{piccoli2006traffic}.

The difficulty of solving the LWR PDE or HJ PDE on the network comes from  the coupling nature of the PDEs at the junction. Given the initial condition on each link and the network boundary conditions, solving the HJ PDE~\eqref{e:hjpde} on each link requires the internal boundary condition for the entire time domain, which is unknown unless the junction model is solved. On the other hand, solving the junction model at any point in time requires the knowledge of the current local traffic condition on connecting links which comes from the solution of the governing HJ PDEs. We refer to~\cite{piccoli2006traffic,imbert2013hamilton} on the integration of junction models in the network PDEs models, as well as the discussion on the well-posedness of the LWR PDE and HJ PDE on a network.

To approximate solutions to the models on networks, several of the numerical schemes for solving the LWR PDE or HJ PDE on a single link have also been extended to the network with an additional treatment at the junction. Like the link schemes, the network schemes discretize each link into cells and time into steps. At the junction, the internal boundary flows are computed using the traffic density for the LWR PDE~\cite{daganzo1995net} or the cumulative number of vehicles for the HJ PDE~\cite{costeseque2014discussion} at the previous time step in the boundary cells of connecting links. After the boundary flow at the next step is obtained, the single step evolution of the traffic state on each link is computed using the link update scheme, such as the Godunov scheme~\cite{godunov1959scheme}. 

Similarly, the single link semi-explicit HJ PDE solver~\cite{claudel2010lax1, claudel2010lax2} can also be extended to the network using a sequential update scheme. Suppose a time grid (not necessarily uniform) with $j_{max}$ intervals is provided at a junction $v$. At each time interval $j$, a junction solver can be formulated as a \textit{convex program} (CP) $\mathcal{CP}_j$ as shown in Section~\ref{subsec:single_step_JS}, which computes the optimal internal boundary flow with respect to the distribution or priority parameters. However, the construction of $\mathcal{CP}_j$ relies on the optimal internal boundary flow solutions up to interval $j-1$. Consequently, computing the internal boundary flows over the entire time horizon requires solving the sequence of convex programs $\mathcal{CP}_j, \forall j \in \mathcal{J}$ consecutively.

Alternatively, this article develops a numerical scheme which  reformulates the sequence of convex programs $\mathcal{CP}_j,\forall j\in \{1,2,\ldots, j_{max}\}$ as an equivalent single convex program.
We show the dependency of $\mathcal{CP}_{j}$ on the optimal solutions up to interval $j-1$ can be relaxed if the objective function of the equivalent single convex program is properly constructed. 
The constraint set of the single convex program over the entire time horizon is simply the union of the constraint sets of the sequence of convex programs with the optimal internal boundary flow solutions at each interval substituted by the corresponding decision variables. Then, the objective function is designed to guarantee the equivalence of the single convex program to the sequence of convex programs $\mathcal{CP}_j, \forall j \in \{1, 2, \ldots, j_{max}\}$. Finally, the obtained internal boundary flows are used as the boundary conditions to solve the corresponding HJ PDE on each link using the semi-explicit single link HJ PDE solver~\cite{claudel2010lax1, claudel2010lax2}. 

The main contribution of this article is the development of the single convex program scheme for computing the internal boundary flows at a merge or diverge junction. Compared to the discretization based methods~\cite{costeseque2014discussion,daganzo1995net,  godunov1959scheme}, the proposed convex optimization scheme does not require discretization of the time-space domain except at the initial time, and at the link spatial boundaries. Moreover, it provides a natural framework for optimal traffic control applications as demonstrated in an example. 

Note that there are other related approaches that also do not require discretization of the time-space domain, including our earlier result on optimal traffic control on networks~\cite{yanningControl} and a recent continuous-time solver of traffic dynamics on the network~\cite{han2015continuous}. Our earlier work~\cite{yanningControl} investigates control of the HJ PDE on a network and assumes all junctions are fully signalized by traffic actuators. Therefore, it does not require a model of the traffic dynamics at the junction and consequently it cannot be used to solve the HJ PDE on a network when the junction dynamics are prescribed. The continuous-time numerical solver~\cite{han2015continuous} for computing the evolution of traffic dynamics on a network uses a link-based kinematic wave model for the link and a mixed integer optimization program for solving the junction problem. In contrast, this article formulates a single convex program to solve  the HJ PDE on the network over the entire time horizon.

The remainder of the article is organized as follows. Section~\ref{s:link_model} reviews the semi-explicit HJ PDE solver on a single link, which enables explicit formulation of the upper bound of the internal boundary flows that can be sent or received on each link at any time. These upper bounds define the feasible set of the convex program for computing the internal boundary flows. In Section~\ref{s:junc_model}, we first describe the behavioral models of the merge and the diverge junction used in this article. Then we show at any time interval, the selected junction models can be posed as a convex program. Section~\ref{subsec:all_step_JS} presents our main contribution, where we formulate a single convex program for solving the junction model over the entire time horizon.
Finally in Section~\ref{sec:on_ramp_control}, an on-ramp metering controller which improves the safety at a work zone by alleviating congestion is proposed to demonstrate the potential of the framework.

\section{Sending and receiving boundary flows on a single link}
\label{s:link_model}

This section first reviews the semi-explicit HJ PDE solver on a single link~\cite{claudel2010lax1}~\cite{claudel2010lax2}. Given the initial condition, the upstream and downstream boundary conditions, the HJ PDE modeling the traffic dynamics on a single link can be semi-explicitly solved. Based on this semi-explicit HJ PDE framework, we then show that if the upstream or downstream boundary condition is unknown, then the upper bound for the boundary flow can be obtained, which denotes the maximum traffic flow that can be sent or received on a link. The obtained bounds are equivalent to the maximum supply and demand~\cite{lebacque1996godunov}, but it can be computed without discretizing the time and space domain as required in the cell transmission model~\cite{daganzo1994cell}. The upper bounds are later used to compute the internal boundary flow solution to the junction models. This section discusses the formulation of the upper bounds for a  single link $l$. The link ID subscript $l$ is included in the notation in this section indicating that the same formulation will later be applied to each link in a network.

\subsection{Semi-explicit HJ PDE solver on a single link}
This subsection reviews the semi-explicit HJ PDE solver~\cite{claudel2010lax1, claudel2010lax2, mazare2011analytical}. 
In the remainder of this article, we further assume the Hamiltonian $\psi_{l}(\cdot)$ on the link is defined as a piecewise affine function~\cite{daganzo1994cell}~\cite{Daganzo05}:
\begin{equation}
\label{e:hamiltonian}
\psi_{l}(\rho) = \left\{
     \begin{array}{ll}
       v^{f}_{l}  \rho & \text{if} \;\; \rho \in \left[0,\rho_{l}^{\kappa} \right],\\
       w_{l} (\rho-\rho_{l}^{m}) & \text{if} \;\; \rho \in \left[\rho_{l}^{\kappa}, \rho_{l}^{m}\right].
     \end{array}
   \right.
\end{equation}
The parameters $\rho_{l}^{\kappa}$, $\rho_{l}^{m}$, $v_{l}^{f}$, and $w_{l}$ represent the critical density, the maximal density, the free flow speed, and the maximum negative congestion wave speed on link $l$. The capacity is then computed as $q_{l}^{max} = v^{f}_{l}\rho^{\kappa}_{l}$.
These parameters are assumed to be known, and can be obtained either from the Highway Capacity Manual~\cite{manual2000highway} or calibrated from measurement data~\cite{horowitz2009automatic}. \edit{Besides the triangular fundamental diagram, the proposed convex scheme can be applied to all concave piecewise linear fundamental diagram, which allows the formulation of piecewise linear constraints. Other concave fundamental diagrams (e.g., the quadratic Greenshields diagram) can be approximated by concave piecewise linear functions for the convex scheme to be applicable.}

In the semi-explicit HJ PDE solver, the initial and boundary conditions of the HJ PDE~\eqref{e:hjpde} on the link $l$ are given by piecewise affine functions defined on an arbitrarily discretized grid. Note that the discretized grid is only required at the boundary of the time-space domain $\left\lbrace \{0\}\times [a_{l}, b_{l}] \right\rbrace \;  $ $\cup \; \left\lbrace[0, t_{max}] \times \lbrace a_{l}, b_{l}\rbrace \right\rbrace$, which is fundamentally different from the discretization of the entire time-space domain $[0, t_{max}] \times [a_{l}, b_{l}]$ into cells and steps in other schemes~\cite{daganzo1994cell, godunov1959scheme}. Specifically, the initial condition at $t=0$ is defined over an arbitrary space grid $\lbrace x_{0}, x_{i}, \forall i \in \mathcal{I}:= \{1, 2, \cdots, i_{max}\} \mid  x_{0}= a_{l}, x_{i} = a_{l} + \sum_{\eta=1}^{i}\Delta x_{\eta}\rbrace$, where $\Delta x_{\eta}$ is the length of spatial interval $\eta$. Similarly, the upstream $x=a_{l}$ and downstream $x=b_{l}$ boundary conditions are defined over a time grid $\lbrace t_{0}, t_{j}, \forall j \in \mathcal{J} := \{1, 2, \cdots, j_{max}\}  \mid t_{0} = 0, t_{j} = \sum_{\eta=1}^{j}\Delta t_{\eta}\rbrace$, where $\Delta t_{\eta}$ is the duration of the temporal interval $\eta$. For conciseness of notation, we omit the subscript $l$ for the grids, which may be link specific in the general case.

On the time space grid $\{ t_{0}, t_{1}, \ldots, t_{j_{max}}\} \times \{ x_{0}, x_{1}, \ldots, x_{i_{max}} \}$, the initial and boundary conditions for the HJ PDE~(\ref{e:hjpde}) are defined as piecewise affine functions, \edit{which are piecewise linear in closed intervals}:
\begin{equation}\label{e:hj_icbc}
\begin{array}{cl}
\mathbf{M}_{l,0}(x) &= \left\lbrace c_{l,0}^{i}(x)  \text{ if } x \in [x_{i-1}, x_{i}] \mid i\in \mathcal{I} \right\rbrace,\\
\mathbf{M}_{a_{l}}(t) &= \left\lbrace c_{a_{l}}^{j}(t) \text{ if } t \in [t_{j-1}, t_{j}] \mid j\in \mathcal{J} \right\rbrace,\\
\mathbf{M}_{b_{l}}(t) &= \left\lbrace c_{b_{l}}^{j}(t) \text{ if } t \in [t_{j-1}, t_{j}] \mid j\in \mathcal{J} \right\rbrace.
\end{array}
\end{equation}

The terms $c_{l,0}^{i}(x), c_{a_{l}}^{j}(t), c_{b_{l}}^{j}(t)$ respectively represent the affine initial or boundary condition defined in the $i$-$th$ space or $j$-$th$ time interval. For compactness, we denote the set of affine initial and boundary conditions for all intervals by $\mathcal{C}_{l}$:
\[
\mathcal{C}_{l} := \lbrace c^{i}_{l, 0}(x), \, c^{j}_{a_{l}}(t), \, c_{b_{l}}^{j}(t) \mid \forall i \in \mathcal{I}, j \in \mathcal{J} \rbrace 
\]

In general, the Moskowitz solution $\mathbf{M}_{l}(t,x)$ to the HJ PDE~(\ref{e:hjpde}) cannot be computed explicitly for arbitrary piecewise affine initial and boundary conditions~\eqref{e:hj_icbc}. However, each affine initial and boundary condition defined in their respective interval, e.g., $c_{l,0}^{i}, c_{a_{l}}^{j}, c_{b_{l}}^{j}$, can be used to compute an explicit \textit{partial solution} in the time-space domain $[0, t_{max}]\times [a_{l}, b_{l}]$ by the \textit{Lax-Hopf} formula. At each point $(t,x)$, a partial solution for each affine initial or boundary condition in $\mathcal{C}_{l}$ can be obtained. The main result of the single link HJ PDE solver~\cite{claudel2010lax1, claudel2010lax2, mazare2011analytical} shows that the Moskowitz solution is the minimum of all partial solutions at $(t,x)$, which is known as the \textit{inf-morphism} property. Using the \textit{Lax-Hopf} formula and  the \textit{inf-morphism} property, the Moskowitz solution $\mathbf{M}_{l}(t,x)$ in the domain $[0,t_{max}]\times [a_{l}, b_{l}]$ can be computed semi-explicitly as shown next.

\begin{proposition}
\label{prop:LaxHopf}
[\textbf{Explicit partial solution}~\cite{aubin2008dirichlet, claudel2010lax1, claudel2010lax2} ] The partial solution, i.e., $\mathbf{M}^{c}_{l}(t,x)$ in the domain $[0,t_{max}]\times[a_{l}, b_{l}]$ associated with each affine initial or boundary condition $c \in \mathcal{C}_{l}$~\eqref{e:hj_icbc} can be explicitly expressed as a linear function of the initial and boundary conditions using the Lax-Hopf formula.

The partial solution $\mathbf{M}_{l}^{c^{j}_{a_{l}}}(t,x)$ associated with the upstream boundary condition $c^{j}_{a_{l}}(t), \,\forall j \in \mathcal{J}$ is written as:
{\footnotesize
\begin{equation}\label{e:LaxHopf}
\mathbf{M}_{l}^{c^{j}_{a_{l}}}(t,x)= \left\lbrace
\begin{array}{ll}
c_{a_{l}}^{j}(t_{j-1}) + \left(\frac{c_{a_{l}}^{j}(t_{j}) - c_{a_{l}}^{j}(t_{j-1})}{\Delta t_{j}}\right) \left(t-\frac{x-a_{l}}{{v_{l}^f}}-t_{j-1}\right), & {\rm if } \;\;t_{j-1}+\frac{x-a_{l}}{{v^{f}_l}}\leq t \\
&{\rm and}\; t< t_{j}+\frac{x-a_{l}}{{v^f_l}},\\
c_{a_{l}}^{j}(t_{j})+\rho_{l}^{\kappa} {v_{l}^f}\left(t-t_{j}-\frac{x-a_{l}}{{v_{l}^f}}\right), &{\rm if } \;\; t \geq t_{j}+\frac{x-a_{l}}{{v_{l}^f}},\\
+\infty  &{\rm otherwise}. 
\end{array}
\right.
\end{equation}
}
The explicit solutions $\mathbf{M}_{l}^{c_{l,0}^{i}}(t,x)$ and $\mathbf{M}_{l}^{c_{b_{l}}^{j}}(t,x)$ associated with the initial and downstream boundary conditions are defined similarly, see~\cite{mazare2011analytical} for a complete description.
\end{proposition}

The partial solution domain (i.e., where $\mathbf{M}_{l}^{c_{a_{l}}^{j}}(t,x)$ is finite) for each affine initial or boundary condition $c \in \mathcal{C}_{l}$ consists of two parts, namely the characteristic domain and the fan domain. In the partial solution associated with the upstream boundary conditions~\eqref{e:LaxHopf}, the characteristic domain is {\footnotesize $\left\lbrace (t,x) \mid t_{j-1} + (x-a_{l})/v^{f}_{l} \leq t < t_{j}\right.$ $\left.+ (x-a_{l})/v^{f}_{l} \right\rbrace$} in the first line of~\eqref{e:LaxHopf} and the fan domain is {\footnotesize $\left\lbrace (t,x) \mid t \geq t_{j} + (x-a_{l})/v^{f}_{l} \right\rbrace$ } in the second line. The vehicle speed and density are constant in the characteristic domain, while the fan domain represents a rarefaction wave connecting the to the characteristic domain of adjacent affine initial or boundary conditions. We refer to~\cite{mazare2011analytical} for a detailed interpretation of
the partial solutions. Physically, the partial solution gives the largest possible vehicle ID in the solution domain by only considering the information in each affine initial or boundary condition.

By the explicit formula~\eqref{e:LaxHopf}, a partial solution $\textbf{M}^{c}_{l}(t,x)$ can be computed for each $c \in \mathcal{C}_{l}$ at each point $(t,x)$. For the Moskowitz solution at $(t,x)$ to be compatible with all affine initial and boundary conditions, it must be less than or equal to the smallest vehicle ID computed by all partial solutions. The following proposition constructs the solution to HJ PDE~(\ref{e:hjpde}) from the set of partial solutions.

\begin{proposition}
\label{prop:inf_morphism}
[\textbf{Inf-morphism property}~\cite{aubin2008dirichlet, claudel2010lax1, claudel2010lax2}]
The Moskowitz solution $\mathbf{M}_{l}(t,x)$ to the HJ PDE~\eqref{e:hjpde} with piecewise affine initial and boundary conditions~\eqref{e:hj_icbc} can be computed as the minimum of all partial solutions defined in Proposition~\ref{prop:LaxHopf} associated with each affine initial and boundary condition:
\[
{\bf M}_{l}(t,x)= {\min_{c \in  \mathcal{C}_{l}  } {\bf M}_{l}^{c}(t,x)  }, \;\; \forall (t,x) \in \left[0,t_{\max}\right] \times \left[a_{l}, b_{l}\right].
\]
\end{proposition}

In summary, the traffic density on a single link can be computed as follows: (\emph{\rmnum 1}) compute the partial solutions $\mathbf{M}_{l}^{c}(t,x)$, $\forall c \in \mathcal{C}$; (\emph{\rmnum 2}) compute the minimum among the set of partial solutions at $(t,x)$ to obtain the Moskowitz solution $\mathbf{M}_l(t,x)$; (\emph{\rmnum 3}) take the derivative of $\mathbf{M}_{l}(t,x)$ with respect to $x$ to recover the traffic density $\rho_{l}(t,x)$.

\subsection{Linear constraints on the boundary flows}
The semi-explicit single link HJ PDE solver assumes the initial and boundary conditions of the link are given. In the cases when the downstream or the upstream boundary condition is unknown, a feasible set can be computed denoting the maximum flow that can be sent or received on the link based on the initial condition and the boundary condition at the other end of the link.

By Proposition~\ref{prop:compatibility_condition}, any boundary flow value in the feasible set can be prescribed as a strong boundary condition for the link while guaranteeing the existence of a unique weak solution to the HJ PDE.

\begin{proposition}
\label{prop:compatibility_condition}
[\textbf{Compatibility conditions}~\cite{claudel2011convex}]
Suppose the initial and the downstream boundary conditions are given in a piecewise affine form for the HJ PDE for a link $l$, the upstream boundary flow data $\tilde{q}_{a_{l}}(t)$ in the continuous time domain prescribes the boundary condition to the HJ PDE~\eqref{e:hjpde} in the strong sense, $a.e.\, t\in (0, t_{max}], \mathbf{M}_{l}(t,a_{l}) =  \mathbf{M}_{a_{l}}(t) = \int_{\tau = 0}^{t} \tilde{q}_{a_{l}}(\tau) {\rm d}\tau$, if and only if:
\begin{equation}\label{e:compatibility_condition_1}
 \begin{array}{lll}
\mathbf{M}_{a_{l}}(t) \leq \underset{\forall c \in \mathcal{C}_{l}}{\operatorname{min}}\,  \mathbf{M}_{l}^{c}\left( t, a_{l} \right), &\quad \forall t \in [0, t_{max}].
\end{array}
\end{equation}
Similarly, given piecewise affine initial and upstream boundary conditions, the downstream boundary flow data $\tilde{q}_{b_{l}}(t)$ in the continuous time domain prescribes the boundary condition to the HJ PDE~\eqref{e:hjpde} in the strong sense, $a.e.\, t\in (0, t_{max}], \mathbf{M}_{l}(t,b_{l}) = \mathbf{M}_{b_{l}}(t) =  \int_{\tau = 0}^{t} \tilde{q}_{b_{l}}(\tau) {\rm d}\tau + c_{l,0}^{i_{max}}(b_{l}) $, if and only if:
\begin{equation}\label{e:compatibility_condition_2}
\begin{array}{lll}
\mathbf{M}_{b_{l}}(t) \leq \underset{\forall c \in \mathcal{C}_{l}}{\operatorname{min}}\,  \mathbf{M}_{l}^{c}\left( t, b_{l} \right), &\quad \forall t \in [0, t_{max}].
\end{array}
\end{equation}
\end{proposition}
The magnitude of the term $c_{l,0}^{i_{max}}(b_{l})$ gives the number of vehicles initially on the link.

The \textit{compatibility conditions} give the upper bound of the boundary flows in a continuous functional space in which it is difficult to be used for analyzing the junction dynamics. In addition, as shown in~\cite{claudel2010lax1}~\cite{claudel2010lax2}, if the given initial and boundary conditions are piecewise affine and the fundamental diagram is triangular, then the unknown boundary flow belongs to a piecewise constant functional space. Therefore, in the numerical implementation, we assume an arbitrary boundary grid with interval length $\Delta t_{j}, j\in \mathcal{J}$ is provided. At each interval $j$, the continuous boundary flow data $\tilde{q}_{a_{l}}(t)$ is approximated by the average flow, i.e., $q_{a_{l}}(j) = \frac{1}{\Delta t_{j}} \int_{t_{j-1}}^{t_{j}} \tilde{q}_{a_{l}}(\tau) {\rm d}\tau, t \in [t_{j-1},\, t_{j}]$ at the upstream boundary. Similarly, $\tilde{q}_{b_{l}}(t)$ is approximated by ${q}_{b_{l}}(j) = \frac{1}{\Delta t_{j}} \int_{t_{j-1}}^{t_{j}} \tilde{q}_{b_{l}}(\tau) {\rm d}\tau,\, t \in [t_{j-1}, t_{j}]$ at the downstream boundary. This approximation allows the construction of an explicit form of the constraints for the boundary flows which is essential for solving the junction models. 

By applying the \textit{compatibility conditions} at the boundary grid points $(t,x) \in \lbrace t_{0}, t_{1},\ldots, t_{j_{max}} \rbrace \times \{a_{l}, b_{l}\}$, the explicit feasible set of boundary flows that can be sent $\mathcal{F}_{l, s}$ or received $\mathcal{F}_{l, r}$ on the link can be obtained. Recall the relationship between the the Moskowitz downstream boundary condition and the boundary flow, i.e., $\mathbf{M}_{b_{l}}(t_{j}) = \sum_{\eta=1}^{j} q_{b_{l}}(\eta)\Delta t_{\eta} + c_{l,0}^{i_{max}}(b_{l})$. Given the initial condition $c_{l,0}^{i}, \forall i \in \mathcal{I}$ and the upstream boundary condition $c_{a_{l}}^{j}, \forall j \in \mathcal{J}$, the feasible set of downstream boundary flows that can be sent on the link at each interval is defined as,
{\footnotesize
\begin{equation} \label{e:feasible_region_1}
\mathcal{F}_{l, s} := \left\lbrace  q(j), \forall j \in \mathcal{J} \mid
\sum_{\eta=1}^{j} q(\eta)\Delta t_{\eta} + c_{l,0}^{i_{max}}(b_{l}) \leq \mathbf{M}_{l}^{c}( t_{j}, b_{l}) , \forall j\in \mathcal{J}, \forall c \in \mathcal{C}_{l} \right\rbrace,
\end{equation}
}
where the subscript $s$ denotes the sending flow and $l$ is the link label.
Similarly, given the initial condition $c_{l,0}^{i}, \forall i \in \mathcal{I}$ and the downstream boundary condition $c_{b_{l}}^{j}, \forall j \in \mathcal{J}$, the explicit feasible set $\mathcal{F}_{l, r}$ of the upstream boundary flows that can be received on the link is formulated as follows,
{\footnotesize
\begin{equation} \label{e:feasible_region_2}
\mathcal{F}_{l, r} := \left\lbrace q(j), \forall j \in \mathcal{J} \mid
\sum_{\eta=1}^{j} q(\eta)\Delta t_{\eta} \leq {\mathbf{M}}_{l}^{c}( t_{j}, a_{l} ) , \forall j\in \mathcal{J}, \forall c \in \mathcal{C}_{l}
\right\rbrace,
\end{equation}
}
where the subscripts $l, r$ denote the receiving flow on link $l$. It is easy to verify that the inequality constraints are linear in the unknown boundary flows $q(j)$ by using the explicit forms of $\mathbf{M}_{l}^{c}(t,x)$ in~\eqref{e:LaxHopf}, and realizing the relationship $q(j)\Delta t_{j} = c_{a_{l}}^{j}(t_{j}) - c_{a_{l}}^{j}(t_{j-1})$ for the downstream boundary flow~\eqref{e:feasible_region_1}, and $q(j)\Delta t_{j} = c_{b_{l}}^{j}(t_{j}) - c_{b_{l}}^{j}(t_{j-1})$ for the upstream boundary flow~\eqref{e:feasible_region_2}. Moreover, the capacity constraints, i.e., $q(j) \leq q_{l}^{max}$ are built into the feasible set by the constraints $\mathbf{M}_{a_{l}}(t_{j}) \leq \mathbf{M}_{l}^{c_{a_{l}}^{j-1}}(t_{j}, a_{l}),\; \mathbf{M}_{b_{l}}(t_{j}) \leq \mathbf{M}_{l}^{c_{b_{l}}^{j-1}}(t_{j}, b_{l})$.

In the extension of the single link HJ PDE solver to a network, the feasible sets of the sending and receiving flows of links are used to compute the internal boundary flows based on the junction models which is discussed in detail in the next section.

\section{Convex formulation of traffic on networks}\label{s:junc_model} 
\begin{figure}[h]
 	\begin{subfigure}{0.5\textwidth}
        \centering
        \includegraphics[width=0.9\textwidth]{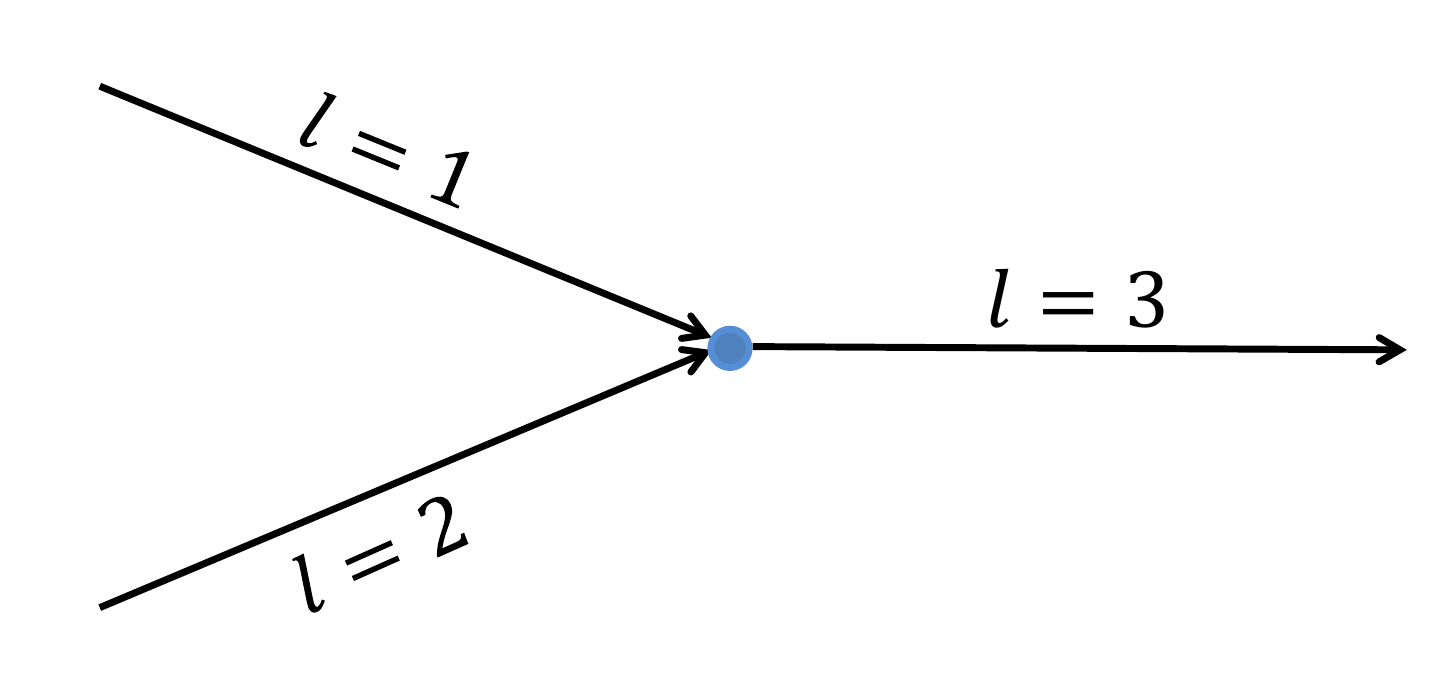}
        \caption{Merge}
        \label{fig:merge}
    \end{subfigure} 
    \begin{subfigure}{0.47\textwidth}
        \centering
        \includegraphics[width=0.9\textwidth]{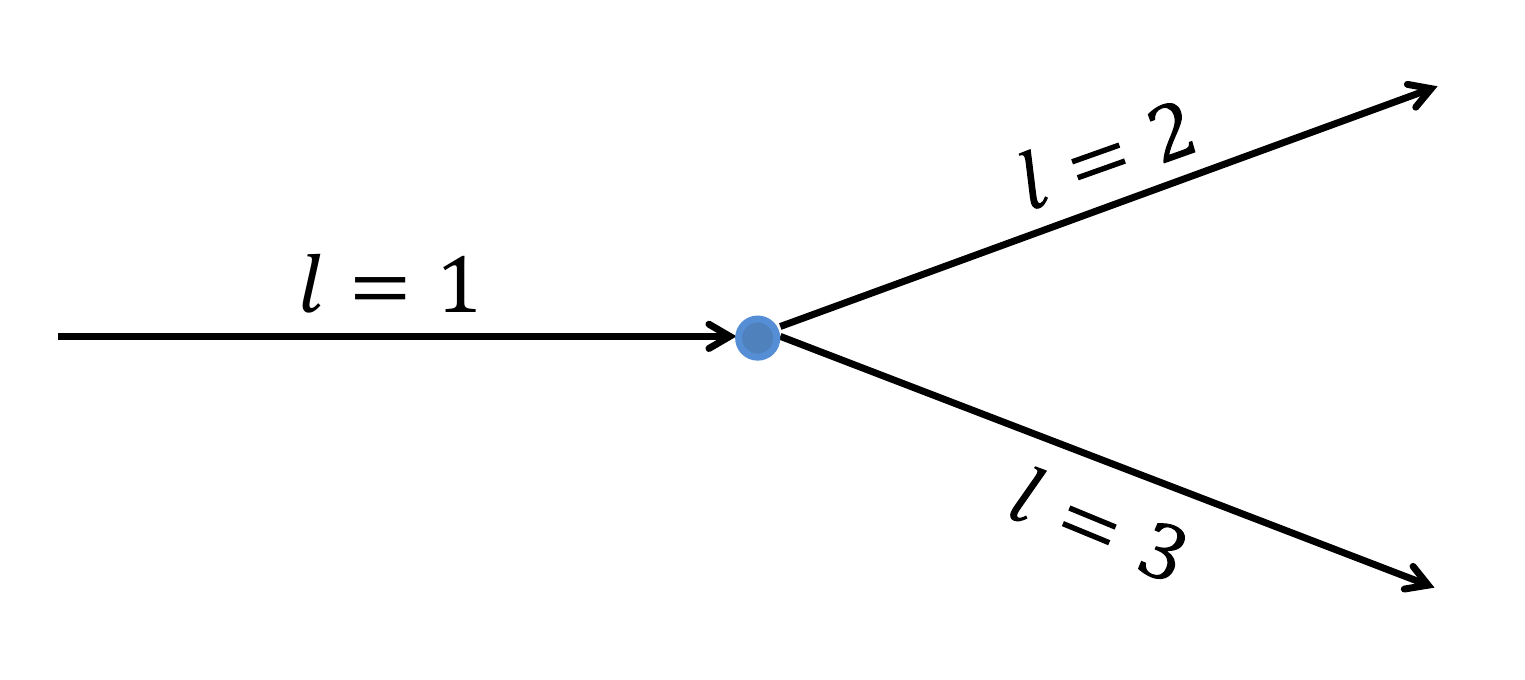}
        \caption{Diverge}
        \label{fig:diverge}
    \end{subfigure} 
    \caption{A transportation network containing a merge or diverge junction with links indexed by $l$.}
    \label{fig:network}
\end{figure}

This section focuses on junction models and the development of a junction solver which computes the internal boundary flows on the network. Since the emphasis of this article is the formulation of a convex program for solving the selected junction models, we focus on a network consisting of three links connected by a merge or diverge junction as shown in Fig.~\ref{fig:network} in the remainder of this article. \edit{In addition to the merge and diverge, a simpler junction is the connection where one upstream link is connected to a downstream link. The connection junction is useful in modeling the road network when the physical property of the road changes (e.g., reduction of lanes).}

\subsection{Junction models on a network}\label{subsec:junction_models}
A junction model describes how the internal boundary flows are distributed across the junction. Specifically, a junction model defines a unique internal boundary flow solution which reflects realistic physical behavior of traffic, such as flow maximization and routing preferences.

\textbf{Merge Model}. This article adopts an existing model~\cite{piccoli2006traffic, han2015continuous} for the merge junction in Fig.~\ref{fig:merge}. \edit{The merge model can be summarized into the following three rules:
\begin{enumerate}[label={\textbf{(A\arabic*)}}, ref=(A\arabic*)]
\item \label{r:1_merge} The mass across the junction is conserved.
\item \label{r:2_merge} The throughput flow is maximized subject to the maximum flow that can be sent or received on each connecting link.
\item \label{r:3_merge}  The distribution of the internal boundary flows, i.e., $q_{a_{3}} \mapsto (q_{b_{1}}, q_{b_{2}})$, satisfies a priority equation $q_{b_{2}} = Pq_{b_{1}}$, where $P$ is a prescribed parameter that models the priority of upstream flows. When \ref{r:3_merge} conflicts with \ref{r:2_merge}, that is, the internal boundary flow solution that satisfies the priority equation does not maximize the throughput, then \ref{r:3_merge} is relaxed, i.e., the solution satisfies \ref{r:2_merge} and minimizes the deviation from the prescribed priority parameter, e.g.,  $\| q_{b_{2}}/q_{b_{1}} - P\|_{1}$.
\end{enumerate}
}

\textbf{Diverge Model}. At a diverge junction in Fig.~\ref{fig:diverge}, this article proposes a model defined by the following rules.
\begin{enumerate}[label={\textbf{(A\arabic*')}}, ref=(A\arabic*')]
\item \label{r:1} The mass across the junction is conserved.
\item \label{r:2} The throughput flow is maximized subject to the maximum flow that can be sent or received on each connecting link.
\item \label{r:3}  The distribution of the internal boundary flows, i.e., $q_{b_{1}} \mapsto (q_{a_{2}}, q_{a_{3}})$, satisfies $q_{a_{3}} = Dq_{a_{2}}$, where $D$ is a prescribed parameter that models the routing preference to the downstream links. When \ref{r:3} conflicts with \ref{r:2}, that is, the internal boundary flow solution that satisfies the distribution equation does not maximize the throughput, then \ref{r:3} is relaxed, such that the solution satisfies \ref{r:2} and minimizes the deviation from the prescribed distribution parameter, e.g.,  $\| q_{a_{3}}/q_{a_{2}} - D\|_{1}$.
\end{enumerate}

The proposed diverge model is a \edit{FIFO model with a varying distribution parameter}. The {classic FIFO diverge model maximizes the throughput subject to the distribution rule $q_{a_{3}}(j) = Dq_{a_{2}(j)}$ with a constant distribution parameter $D$}. The \edit{classic} FIFO model circumvents the difficulty of resolving the conflicts between the throughput maximization and flow distribution, but it produces unrealistic solutions in some applications. For example, using the \edit{classic} FIFO model, a blocked offramp will completely stop the traffic on all lanes of a multi-lane highway, which is unlikely. To resolve this issue, several diverge junction models were proposed previously, such as a multi-lane junction model \edit{(non-FIFO)}~\cite{herty2003modeling}, a dynamic distribution parameter~\cite{jin2003distribution}, and a junction model with internal dynamics~\cite{lebacque2005intersection}. In the same spirit of these models, this article proposes a diverge junction model that produces similar traffic condition dependent solutions without introducing additional complexity \edit{of non-FIFO models} on the traffic dynamics. The main assumption of the proposed diverge model is that drivers will reroute to the other link if the initially desired link becomes congested~\cite{holden1995mathematical}. The rerouting assumption makes the composition of queuing vehicles on the upstream link time-invariant which is critical for the unique solution to be consistent~\cite{lebacque2005first}.


\edit{The connection junction model is significantly simpler compared to the merge and diverge models since there is no distribution or priority parameters involved. Therefore, the connection model simply maximizes the throughput.}

The structure of the merge and diverge models used in this article are similar, i.e., both maximize the throughput and then minimize the deviation from the prescribed priority or distribution parameters. Therefore, the remainder of this section will focus on the formulation of a junction solver for the \edit{merge} model and note the same analysis can be easily transferred to the \edit{diverge}. \edit{The connection junction solver will be briefly discussed considering its simplicity.}

\subsection{Junction solver over a single interval}\label{subsec:single_step_JS}
This subsection proposes a junction solver in the form of a convex program that computes the internal boundary flow solution at a single time interval for the \edit{merge} model.

To compute the unique internal boundary flows \edit{$q_{b_{1}}(j), q_{b_{2}}(j)$} at time interval $j$, we assume that the unique internal boundary flow solutions up to interval $j-1$ are given and denoted by $q^{\ast}_{a_{l}}(\eta), q^{\ast}_{b_{l}}(\eta), \eta	\in \{1,2,\ldots, j-1\}$. It should be noted that this assumption requires the \edit{merge} junction problem to be solved sequentially in time. Accordingly, the convex set~\eqref{e:feasible_region_1} of the internal boundary flows that can be sent on link $l$ for all time interval up to $j$ is reduced to a convex set of internal boundary flows that can be sent at interval $j$:
{
\begin{equation}
\label{e:single_step_feasible_region_1}
\mathcal{F}_{l, s}^{j} := \left\lbrace q(j) \mid
q(j)\Delta t_{j} \leq \mathbf{M}_{l}^{c}(t_j, b_{l}) - \sum_{\eta=1}^{j-1}q_{b_{l}}^{\ast}(\eta)\Delta t_{\eta} - c_{l,0}^{i_{max}}(b_{l}), \forall c \in \mathcal{C}_{l} \right\rbrace.
\end{equation}
}
Similarly, the internal boundary flows that can be received on link $l$ at time interval $j$ are subject to a reduced feasible set of~\eqref{e:feasible_region_2}:
{
\begin{equation}
\label{e:single_step_feasible_region_2}
\mathcal{F}_{l, r}^{j}:= \left\lbrace q(j) \mid
q(j)\Delta t_{j} \leq \mathbf{M}_{l}^{c}(t_{j}, a_{l}) - \sum_{\eta=1}^{j-1}q_{a_{l}}^{\ast}(\eta)\Delta t_{\eta} , \forall c \in \mathcal{C}_{l} \right\rbrace.
\end{equation}
}
The terms $\mathbf{M}_{l}^{c}(t_{j}, b_{l}) - c_{l,0}^{i_{max}}(b_{l})$ and $\mathbf{M}_{l}^{c}(t_{j}, a_{l})$ denote the maximum number of vehicles that can be sent or received during time $(0, t_{j})$. The summations of the given boundary flows $\sum_{\eta=1}^{j-1}q_{b_{l}}^{\ast}(\eta)\Delta t_{\eta}$ and $\sum_{\eta=1}^{j-1}q_{a_{l}}^{\ast}(\eta)\Delta t_{\eta}$ represent the number of vehicles that have been sent or received during $(0, t_{j-1})$. Hence, the right-hand side terms in~\eqref{e:single_step_feasible_region_1} and~\eqref{e:single_step_feasible_region_2} are constants representing the maximum number of vehicles that can be sent or received during interval $j$, i.e., $(t_{j-1}, t_{j})$. 

As shown next, the \edit{merge} junction solver is posed as a convex program with a carefully constructed objective function to accommodate the throughput maximization~\ref{r:2_merge} and the flow \edit{priority}~\ref{r:3_merge} objectives. The equations~\eqref{e:single_step_feasible_region_1} and \eqref{e:single_step_feasible_region_2} combined define the constraint set of the convex program.

\edit{
\begin{definition}
\label{prop:single_step_condition_merge}
[\textbf{Merge junction solver over a single interval}]
The junction solver for computing the internal boundary flow solution $(q_{b_{1}}(j), q_{b_{2}}(j))$ at a merge during interval $j$ is formulated in the form of a convex program as follows:
\begin{equation}
\label{CP:single_step_merge}
\begin{array}{ll}
\underset{q_{1}(j), q_{2}(j)}{{\rm Maximize}} & f\left(q_{1}(j), q_{2}(j)\right)\\
{\rm s.t.} & q_{1}(j) \in \mathcal{F}_{1, s}^{j}\quad~\eqref{e:single_step_feasible_region_1}, \\
& q_{2}(j) \in \mathcal{F}_{2, s}^{j}\quad~\eqref{e:single_step_feasible_region_1},\\
& q_{3}(j) \in \mathcal{F}_{3, r}^{j}\quad~\eqref{e:single_step_feasible_region_2},\\
& q_{3}(j) = q_{1}(j) + q_{2}(j),
\end{array}
\end{equation}
where $f(q_{1}(j),q_{2}(j))$ is a convex function of $q_{1}(j), q_{2}(j)$ and satisfies:
{\small
\begin{subequations}
\label{eq:prop_single_step_merge}
\begin{align}
&\frac{\partial f}{\partial q_{l}(j)} > 0, & &\forall l \in \{1,2\},  \label{eq:prop_single_step_merge_1} \\
&\frac{\partial f}{\partial q_{1}(j)} > \frac{\partial f}{\partial q_{2}(j)}, & &{\rm when }\;\; q_{2}(j) \geq Pq_{1}(j), \label{eq:prop_single_step_merge_2} \\
&\frac{\partial f}{\partial q_{1}(j)} < \frac{\partial f}{\partial q_{2}(j)}, & &{\rm when }\;\; q_{2}(j) < Pq_{1}(j).  \label{eq:prop_single_step_merge_3} 
\end{align}
\end{subequations}
}
\end{definition}
}

The junction solver CP~\eqref{CP:single_step_merge} computes the unique internal boundary flow solution defined by the \edit{merge} junction model \ref{r:1_merge},~\ref{r:2_merge}, and \ref{r:3_merge}, as stated in the following proposition.

\edit{
\begin{proposition}
\label{def:junction_solution_merge}
The merge junction solver CP~\eqref{CP:single_step_merge} computes the unique internal boundary flow solution $q^{\ast}(j) = \left( q^{\ast}_{b_{1}}(j), q^{\ast}_{b_{2}}(j), q^{\ast}_{a_{3}}(j) \right)$ at interval $j$, where $q^{\ast}(j)$ satisfies \ref{r:1_merge},~\ref{r:2_merge},~\ref{r:3_merge}:
\begin{enumerate}[label=(\roman*), ref=(\roman*)]
\item \label{c:1_merge} The internal boundary flows satisfy mass conservation rule~\ref{r:1_merge},
$
q^{\ast}_{a_{3}}(j) = q^{\ast}_{b_{1}}(j) + q^{\ast}_{b_{2}}(j).
$
\item \label{c:2_merge} The throughput flow at the junction is maximized subject to the feasible sets on connecting links~\ref{r:2_merge}, i.e., $q^{\ast}(j) \in \mathcal{Q}_{j} := \underset{q(j) \in \mathcal{F}^{j}_{1,s} \times \mathcal{F}^{j}_{2,  s} \times \mathcal{F}^{j}_{3, r} }{ \operatorname{argmax} } q_{a_{3}}(j)$.
\item  \label{c:3_merge} The deviation from the priority equation is minimized~\ref{r:3_merge}, e.g., $q^{\ast}(j) = \underset{q(j)\in \mathcal{Q}_{j}}{\operatorname{argmin} } $ $\| q_{b_{2}}(j) - Pq_{b_{1}}(j) \|_{1}$.
\end{enumerate}
\end{proposition}
}
\begin{proof}
A sketch of the proof is provided here, and a detailed proof appears in Appendix~\ref{appendix:single}. By construction of the junction solver, the constraint set of the convex program is non-empty, which guarantees the existence of a solution. The property~$\ref{c:1_merge}$ is satisfied since it is explicitly included in the constraint set in CP~\eqref{CP:single_step_merge}. The property~$\ref{c:2_merge}$ is guaranteed by the condition~\eqref{eq:prop_single_step_merge_1} since the gradient with respect to each internal boundary flow is strictly positive. The intuition of conditions~\eqref{eq:prop_single_step_merge_2}~\eqref{eq:prop_single_step_merge_3} is that the gradient of the objective function $f$ points towards the line \edit{$q_{b_{2}}(j) = Pq_{b_{1}}(j)$} from both sides in the feasible set. Consequently, the points closer to the distribution line have a smaller deviation \edit{$\| q_{b_{2}}(j)/q_{b_{1}}(j) - P\|_{1}$} as defined in~$\ref{c:3_merge}$. The uniqueness of the solution is proved in the detailed proof using conditions in~\eqref{eq:prop_single_step_merge}.
\end{proof}

The consistency (equivalently invariance) of the \edit{merge} junction solver can be verified by direct application of the definition of invariance~\cite{lebacque2005first}. \\

\begin{figure}[htpb]
    \begin{subfigure}{0.32\textwidth}
        \centering
        \includegraphics[width=\textwidth]{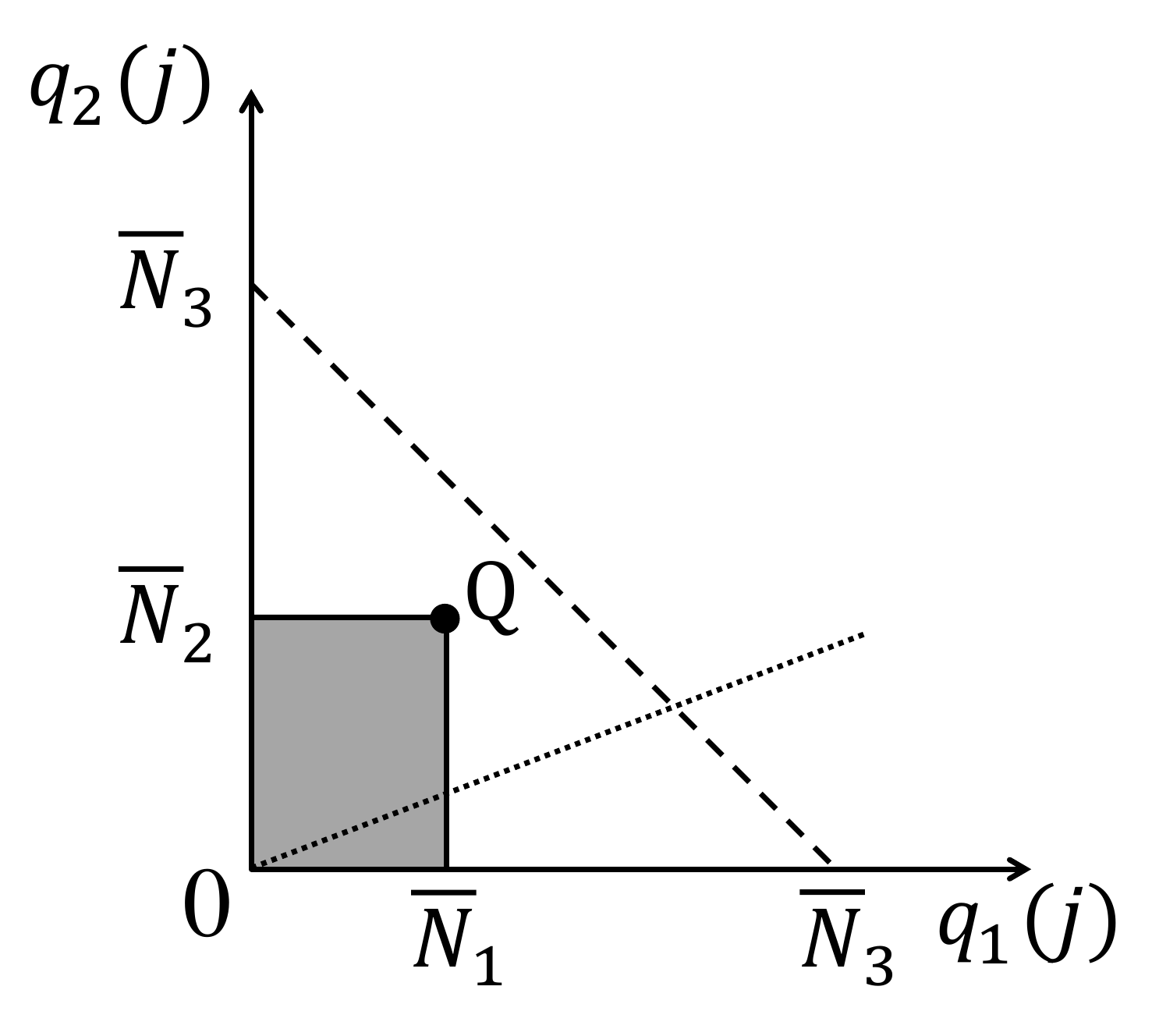}
        \caption{\textbf{Merge Case One}}
        \label{fig:merge_1}
    \end{subfigure}
    \begin{subfigure}{0.32\textwidth}
        \centering
        \includegraphics[width=\textwidth]{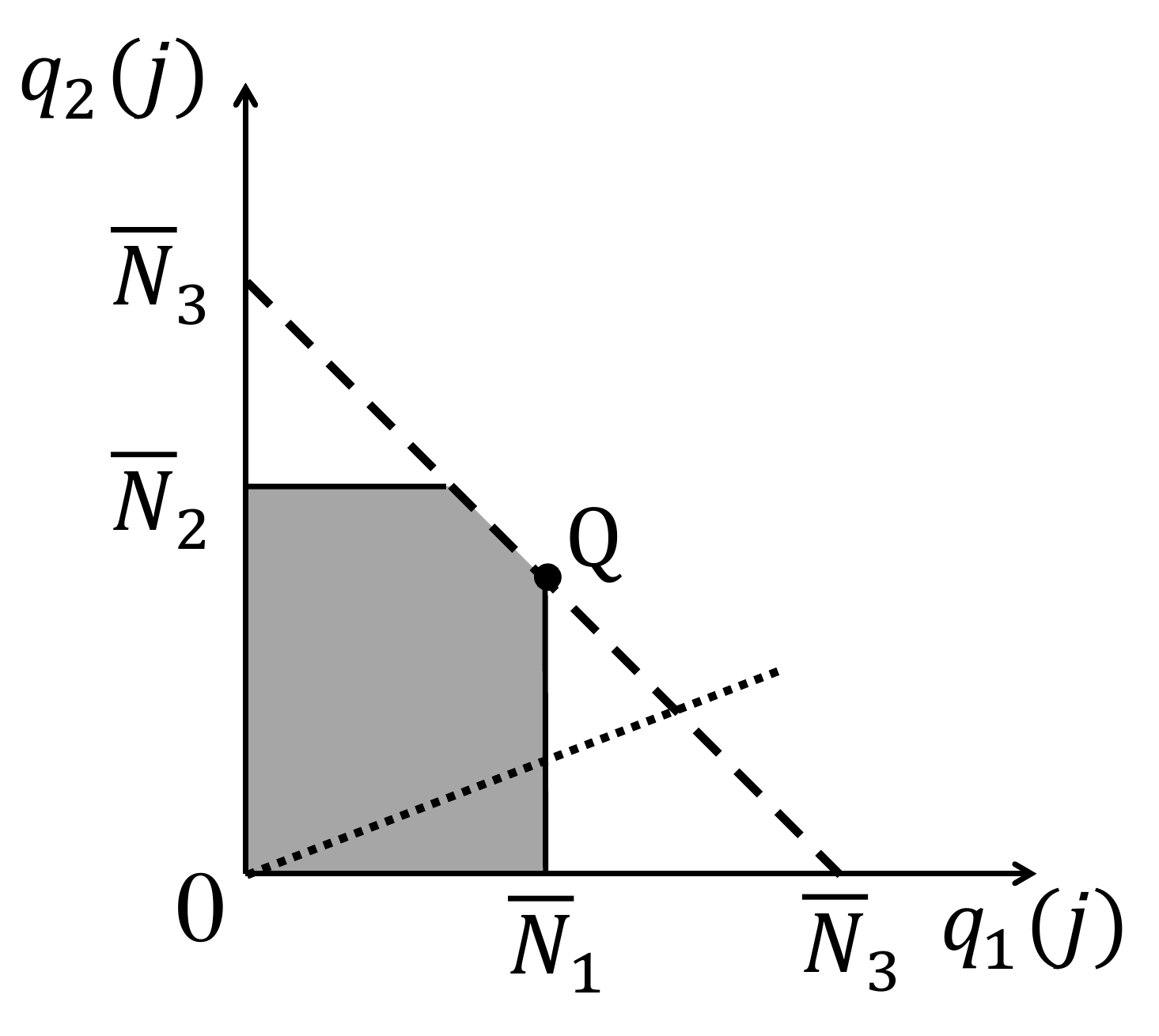}
        \caption{\textbf{Merge Case Two}}
        \label{fig:merge_2}
    \end{subfigure} 
    \begin{subfigure}{0.32\textwidth}
        \centering
        \includegraphics[width=\textwidth]{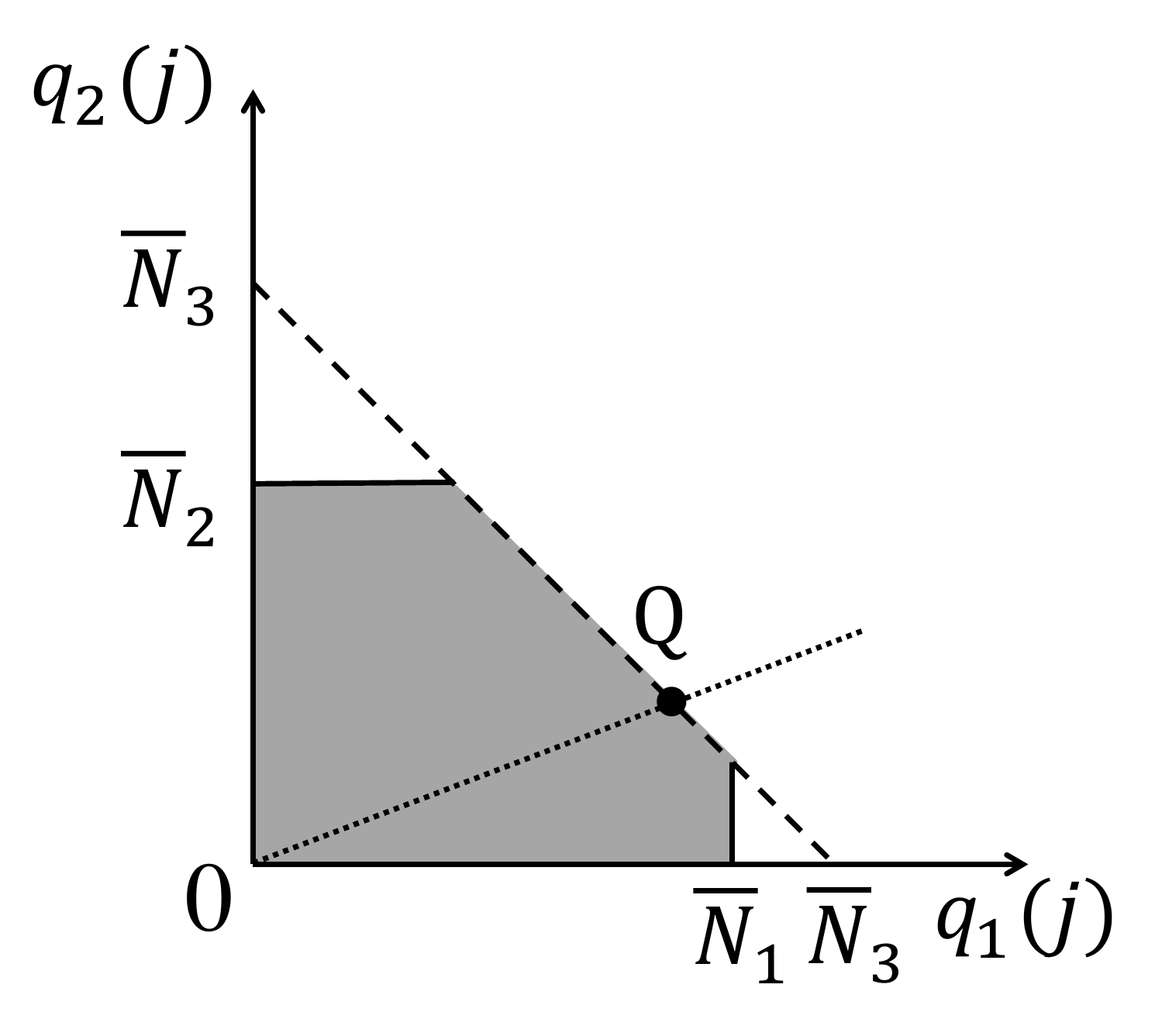}
        \caption{\textbf{Merge Case Three}}
        \label{fig:merge_3}
    \end{subfigure} 
    \caption{\edit{{Three scenarios at a merge.} Link 1 and 2 merge to link 3. The two solid lines represent the maximum sending flow on links 1 and 2. The dashed line denotes the maximum receiving flow on link 3. The dotted line denotes the prescribed priority of sending flows. The shaded area is the feasible set of boundary flows~\eqref{e:single_step_feasible_region_1}~\eqref{e:single_step_feasible_region_2}. The unique solution computed by CP~\eqref{CP:single_step_merge} is depicted by Q.}}
    \label{fig:Merge_scenarios}
\end{figure}

\edit{
The unique solution during a single time interval $j$ computed by the junction solver~\eqref{CP:single_step_merge} for the merge is illustrated in Fig.~\ref{fig:Merge_scenarios}. There are in total three scenarios depending on the feasible sets on links~\eqref{e:single_step_feasible_region_1}~\eqref{e:single_step_feasible_region_2}, namely, \textit{(i)} when the maximum receiving flow on link 3 exceeds the total sending flow from links 1 and 2 in Fig.~\ref{fig:merge_1}; \textit{(ii)} when the maximum receiving flow is smaller than the total sending flow from links 1 and 2 but the prescribed priority ratio can not be followed exactly in Fig.~\ref{fig:merge_2}; \textit{(iii)} when the maximum receiving flow is smaller than the total sending flow on links 1 and 2 and the prescribed priority ratio can be followed exactly in Fig.~\ref{fig:merge_3}. The feasible set of the convex program is denoted by the shaded area, where the upper bounds of the feasible internal boundary flows for three links are respectively  $\bar{N}_{1} = \max \left\lbrace { q_{1}(j) \mid  q_{1}(j)\in \mathcal{F}_{1, s}^{j}} \right\rbrace, \, \bar{N}_{2} = \max \left\lbrace q_{2}(j) \mid q_{2}(j)\in \mathcal{F}_{2, s}^{j}\right\rbrace$, and $\bar{N}_{3} = \max \left\lbrace q_{3}(j) \mid q_{3}(j)\in \mathcal{F}_{3, r}^{j} \right\rbrace$. The solid lines denote the maximum sending flows from links 1 and 2. The dashed line denotes the maximum receiving flow on link 3. The dotted line denotes the prescribed priority of the boundary flows, i.e., $q_{1}(j) = \frac{ q_{3}(j) }{1+P},\; q_{2}(j) = \frac{Pq_{3}(j) }{1+P}$. The unique solution computed by the diverge junction solver CP~\eqref{CP:single_step_merge} is marked at point $Q$. 
}

\edit{
In scenario \textit{(i)}, see Fig.~\ref{fig:merge_1}, the maximum receiving flow on link 3 exceeds the the total flow that can be sent by links 1 and 2 combined. Hence, the single point that maximizes the throughput admits the maximum sending flow from links 1 and 2, and is the optimal solution to CP~\eqref{CP:single_step_merge}.}

\edit{
In scenario \textit{(ii)}, see Fig.~\ref{fig:merge_2}, the total sending flow is higher than the maximum receiving flow on the downstream link, and the sending flows cannot be distributed exactly following the priority rule. In this case, link 3 first admits all flows from the higher priority link (i.e., link 1 in Fig.~\ref{fig:merge_2}), and then admits as much flow as possible for the lower priority link to maximize the throughput. Consequently, the optimal solution to CP~\eqref{CP:single_step_merge} is the solution $Q$ that is closest to the dotted line among the solutions on the dashed line within the feasible set. 
}

\edit{
In last scenario \textit{(iii)}, see Fig.~\ref{fig:merge_3}, the total sending flow from upstream links combined is higher than the maximum receiving flow on link 3, and the sending flow can be distributed exactly following the priority ratio. In this case, there is no conflicts between maximizing the throughput~\ref{r:2_merge} and following the flow priority~\ref{r:3_merge}. Therefore, the solution at $Q$ that satisfies both~\ref{r:2_merge} and~\ref{r:3_merge} is the optimal solution to the merge junction solver CP~\eqref{CP:single_step_merge}.
}

\edit{Similar to the merge model, the diverge model also has three scenarios for determining the unique solution. The three scenarios are depicted and discussed in Appendix~\ref{appendix:diverge}.}

The junction solver CP~\eqref{CP:single_step_merge} applies to an arbitrary convex objective function \edit{$f(q_{1}(j), q_{2}(j))$} that satisfies~\eqref{eq:prop_single_step_merge}, and sub-derivatives can be used if \edit{$f(q_{1}(j), q_{2}(j))$} is not differentiable. Though no explicit form of the objective function is used to define the solver, selecting a suitable objective function is straightforward as shown in the following example.

\edit{
\textbf{Example:}
At a merge, a convex objective function to maximize the throughput~\ref{r:2_merge}, then minimize the deviation from prescribed flow priority ratio~\ref{r:3_merge} for any flow maximizing solution, can be defined as follows:
\begin{equation}\label{e:example1_f}
f(q_{1}(j), q_{2}(j)) = \alpha\left(q_{1}(j)+q_{2}(j)\right)-\beta\left(q_{2}(j)-Pq_{1}(j)\right)^{2},
\end{equation}
where $\alpha$ and $\beta$ are weights defined by:
\[
\alpha = 1- \beta, \;
\beta =\min \left( \frac{1}{1+2P^2q_{1}^{max} + \epsilon},\frac{1}{1+2q_{2}^{max} +\epsilon}\right),\; \text{and } \epsilon = 0.01.
\]
The derivation of the coefficients is presented as follows. Define the objective function as~\eqref{e:example1_f} and assume $\alpha > 0$ and $\beta > 0$.
The conditions~(\ref{eq:prop_single_step_merge_2}) and
(\ref{eq:prop_single_step_merge_3}) in Proposition~\ref{prop:single_step_condition_merge} are trivially satisfied given $\beta > 0$. Then by condition~(\ref{eq:prop_single_step_merge_1}), 
{\small
\[
\begin{array}{ll}
\frac{\partial f}{\partial q_1(j)} = \alpha - 2\beta(q_{2}(j)-Pq_{1}(j))(-P) > 0, &\; \text{and} \; 
\frac{\partial f}{\partial q_2(j)} = \alpha - 2\beta(q_{2}(j)-Pq_{1}(j)) > 0.
\end{array}
\]
}
Let $\alpha = 1- \beta$ and $\beta \in (0,1)$, and rearrange the above inequalities as
\[
\begin{array}{ll}
1 >  \beta\left(1 -2Pq_{2}(j) + 2P^2q_{1}(j) \right), & \quad \text{and} \quad
1 > \beta\left(1+2q_{2}(j)-2Pq_{1}(j)\right).
\end{array}
\]
To guarantee the above inequalities hold for all possible values of $q_{1}(j)$ and $q_{2}(j)$, we use the following chain rule,
\[
\begin{array}{l}
\beta\left(1+2P^2q_{1}^{max}\right) \geq \beta\left(1+2P^2q_{1}(j)\right) \geq \beta\left(1 -2P q_{2}(j) + 2P^2q_{1}(j) \right),\\
\beta\left(1+2q_{2}^{max} \right) \geq \beta\left(1+2q_{2}(j) \right) \geq \beta\left(1+2q_{2}(j)-2Pq_{1}(j)\right).
\end{array}
\]
Hence, it suffices to require
\[
\begin{array}{ll}
1 > \beta\left(1+2P^2q_{1}^{max}\right), & \quad \text{and} \quad 1> \beta\left(1+2q_{2}^{max} \right).
\end{array}
\]
Then the value of $\beta$ can be selected as:
{\small
\[
	\beta = \min \left(\frac{1}{1+2P^2q_{1}^{max} + \epsilon},\frac{1}{1+2q_{2}^{max} +\epsilon }\right) \in [0,1],
\]
}
where $\epsilon$ is a small positive constant (e.g. 0.01) to guarantee the strict inequality. Accordingly, $\alpha$ is computed then by $\alpha = 1 - \beta$.
}

\edit{The connection junction model only requires the mass conservation and maximization of the throughput. Therefore, the connection junction solver can be formulated by maximizing the throughput (i.e., downstream flow of upstream link) subject to the maximum sending and receiving boundary flows defined in~\eqref{e:single_step_feasible_region_1} and \eqref{e:single_step_feasible_region_2}.}

\subsection{Junction solver over the entire horizon} \label{subsec:all_step_JS}
As shown in the previous subsection, the unique solution over a single time interval can be computed by a junction solver in the form of a convex optimization program. When constructing the constraint set of CP~\eqref{CP:single_step_merge} at interval $j$, the unique solutions up to interval $j-1$ were assumed to be known~\eqref{e:single_step_feasible_region_1}~\eqref{e:single_step_feasible_region_2}. As a result, computing the internal boundary flow solution over the entire time horizon requires consecutively constructing and solving a convex program at each time interval. This subsection presents a \edit{merge} junction solver in the form of a single convex program for computing the interval boundary flows over the entire time horizon.

The intuition of the junction solver as a single convex program is that the union of the constraint set for each interval $\mathcal{F}_{l, r}^{j}$ or $\mathcal{F}_{l, s}^{j}$ is a subset of the constraint set $\mathcal{F}_{l, r}$ or $\mathcal{F}_{l, s}$ over the entire time horizon. Given the structure of the constraint set, it is possible to design the objective function for a single convex program such that the solution over the entire time horizon is equivalent to the solution computed by sequentially solving a convex program at each time interval. The following definition articulates the conditions required for the objective function.

\edit{
\begin{definition}
\label{prop:multiple_step_condition}[\textbf{Merge junction solver over the entire time horizon}]
The junction solver for computing the internal boundary flows at a merge $(q_{b_{1}}, q_{b_{2}})$ where $q_{b_{1}} = \{ q_{b_{1}}(j), \forall j \in \mathcal{J} \}$ and $q_{b_{2}} = \{ q_{b_{2}}(j), \forall j \in \mathcal{J} \}$ for the entire time horizon is defined as the following convex program, 
\begin{equation}
\label{CP:multiple_steps_merge}
\begin{array}{cl}
\underset{q_{1}, q_{2}}{\rm Maximize} & f(q_{1}, q_{2}) \\
{\rm s.t.}  & q_{1} \in \mathcal{F}_{1,s}\quad~\eqref{e:feasible_region_1},\\
& q_{2} \in \mathcal{F}_{2,s}\quad~\eqref{e:feasible_region_2},\\
& q_{3} \in \mathcal{F}_{3,r}\quad~\eqref{e:feasible_region_2},\\
& q_{3}(j) = q_{1}(j) + q_{2}(j),\; \forall j \in \mathcal{J},\\
\end{array}
\end{equation}
where $q_{1} = \{q_{1}(j), \forall j \in \mathcal{J}\},\; q_{2} = \{q_{2}(j), \forall j \in \mathcal{J}\},\; q_{3} = \{q_{3}(j), \forall j \in \mathcal{J}\}$. The function $f(q_{1}, q_{2})$ is convex in $q_{1}(j), q_{2}(j),\,\forall j \in \mathcal{J}$ and satisfies:
{\small
\begin{subequations}\label{eq:prop_multiple_step_12}
\begin{align}
&\frac{\partial f}{\partial q_{1}(1)\Delta t_{1}}  > \frac{\partial f}{\partial q_{1}(2)\Delta t_{2}} > \cdots > \frac{\partial f}{\partial q_{1}(j_{max})\Delta t_{j_{max}}} >0, 
\label{eq:prop_multiple_step_merge_1}\\
&\frac{\partial f}{\partial q_{2}(1)\Delta t_{1}}  > \frac{\partial f}{\partial q_{2}(2)\Delta t_{2}}  > \cdots > \frac{\partial f}{\partial q_{2}(j_{max})\Delta t_{j_{max}}} >0,
\label{eq:prop_multiple_step_merge_2}
\end{align}
\end{subequations}
}
and  $\forall j \in \{1, 2, \cdots, j_{max}-1\}$, 
{\small
\begin{subequations}\label{eq:prop_multiple_step_34}
\begin{align}
&\frac{\partial f}{\partial q_{1}(j)\Delta t_{j}} - \frac{\partial f}{\partial q_{2}(j)\Delta t_{j}} - \frac{\partial f}{\partial q_{1}(j+1)\Delta t_{j+1}}>0 \quad {\rm if }\;\; q_{2}(j) \geq Pq_{1}(j),
\label{eq:prop_multiple_step_merge_3}\\
&\frac{\partial f}{\partial q_{2}(j)\Delta t_{j}} - \frac{\partial f}{\partial q_{1}(j)\Delta t_{j}} -\frac{\partial f}{\partial q_{2}(j+1)\Delta t_{j+1}}>0 \quad {\rm if }\;\; q_{2}(j) < Pq_{1}(j),
\label{eq:prop_multiple_step_merge_4}
\end{align}
\end{subequations}
}
and  when $j = j_{max}$, 
{\small
\begin{subequations}\label{eq:prop_multiple_step_56}
\begin{align}
&\frac{\partial f}{\partial q_{1}(j_{max})} - \frac{\partial f}{\partial q_{2}(j_{max})} >0 \quad {\rm if }\;\; q_{2}(j_{max}) \geq Pq_{1}(j_{max}),
\label{eq:prop_multiple_step_merge_5}\\
&\frac{\partial f}{\partial q_{1}(j_{max})} - \frac{\partial f}{\partial q_{2}(j_{max})} <0 \quad {\rm if }\;\; q_{2}(j_{max}) <Pq_{1}(j_{max}).
\label{eq:prop_multiple_step_merge_6}
\end{align}
\end{subequations}
}
\end{definition}
}

Before stating the formal proposition on the equivalence of the junction solver CP~\eqref{CP:multiple_steps_merge} to the single interval junction solver CP~\eqref{CP:single_step_merge} solved sequentially for all intervals, we briefly interpret the conditions~\eqref{eq:prop_multiple_step_12}~\eqref{eq:prop_multiple_step_34}~\eqref{eq:prop_multiple_step_56} on the objective function.

The conditions~\eqref{eq:prop_multiple_step_12} assign higher weights to the internal boundary flows at earlier time intervals, such that the throughput at earlier intervals is first maximized. Consequently, the convex program will produce a solution that satisfies the throughput maximization rule~\ref{r:2_merge} for all intervals. The unique solution also requires minimum deviation of the solution from the prescribed parameter~\ref{r:3_merge}, i.e., \edit{$q_{b_{2}}(j) = Pq_{b_{1}}(j)$} for all intervals $j \in \mathcal{J}$. The conditions~\eqref{eq:prop_multiple_step_34}~\eqref{eq:prop_multiple_step_56} define the direction of the gradient of the objective function as pointing towards the line \edit{$q_{b_{2}}(j) = Pq_{b_{1}}(j)$} from both sides. If \eqref{eq:prop_multiple_step_34}~\eqref{eq:prop_multiple_step_56} are satisfied for CP~\eqref{CP:multiple_steps_merge}, then conditions \eqref{eq:prop_single_step_merge_2}~\eqref{eq:prop_single_step_merge_3} are satisfied for all $CP_{j}$ at all intervals $\forall j \in \mathcal{J}$.

\edit{
\begin{proposition}
\label{prop:multiple_step}
The junction solver~\eqref{CP:multiple_steps_merge} gives the same unique solution $\left\lbrace q_{b1}(j), \right. $ $\left. q_{b_{2}}(j), q_{a_{3}}(j) \mid \forall j \in \mathcal{J} \right\rbrace$ obtained by sequentially solving CP~\eqref{CP:single_step_merge} at each time interval.
\end{proposition}
}
\begin{proof}
A detailed proof is presented in Appendix~\ref{appendix:multiple} and the intuition is briefly described as follows.
The proof of Proposition~\ref{prop:multiple_step} relies on the equivalence of the \textit{Karush-Kuhn-Tucker} (KKT) conditions associated with the sequence of CPs~(\ref{CP:single_step_merge}) and CP~(\ref{CP:multiple_steps_merge}). The main idea is to show the set of KKT multipliers associated with the optimal solution to CP~\eqref{CP:multiple_steps_merge} also satisfies the KKT conditions associated with the same solution to CP~\eqref{CP:single_step_merge} for each time interval. Since the constraints in CP~\eqref{CP:single_step_merge} are linear, the KKT conditions are also sufficient conditions. Therefore, the solution to CP~\eqref{CP:multiple_steps_merge} is also the optimal solution to CP~\eqref{CP:single_step_merge} for each time interval.
\end{proof}

The junction solver CP~\eqref{CP:multiple_steps_merge} does not provide an explicit form of the objective function. Similarly as Proposition~\ref{def:junction_solution_merge}, one can first define a weighted objective function with undetermined weights and then select values for the weights to satisfy the proposed conditions on the objective function. This process is illustrated in the following example.

\textbf{Example:}
For simplicity, suppose the number of time steps $j_{max} = 2$. Define the following objective function for a diverge with weights $\alpha >0, \beta >0, \omega_{1}>0, \omega_{2}>0$:
\edit{
{\small
\[
f(q_{1}, q_{2}) = \sum_{j=1}^{2} w_j\left( \alpha\cdot \left(q_{1}(j) + q_{2}(j) \right) - \beta\cdot \| q_{2}(j) - Pq_{1}(j) \|_{1}\right).
\]
}
}
The term \edit{$q_{1}(j) + q_{2}(j)$} maximizes the throughput at time interval $j$ and the term \edit{$\|  q_{2}(j) - Pq_{1}(j) \|_{1}$} penalizes the deviation from the prescribed \edit{priority} ratio. In addition, weights $w_{j}$ are selected for the two intervals to assign a higher weights to the internal boundary flows at earlier intervals, so the vehicles are not stopped and sent at later time intervals which is unrealistic at a junction without actuators. By applying the conditions~\eqref{eq:prop_multiple_step_12}~\eqref{eq:prop_multiple_step_34}~\eqref{eq:prop_multiple_step_56} on the objective function $f(q_{2}, q_{3})$, the parameters and weight coefficients can be set as follows:
\edit{
{\small
\[
\alpha = 1 + P + {\rm max}(1, P),\quad
\beta = 1, \quad  
\omega_{1} = \frac{\alpha + {\rm max}(1, P)}{1+P} + \epsilon, \quad \omega_{2} = 1, 
\]
}
}
where $\epsilon >0$ is a small positive constant to guarantee the strict inequality. \\

\edit{At the connection junction, the throughput must be maximized at all time intervals. The following lemma shows the unique solution over the entire time horizon for a connection can be computed in a single convex program.}
\edit{
\begin{lemma}
\label{lemma:connection}
Consider a junction in which one upstream link (link 1) connects to one downstream link (link 2). The unique boundary flow solution $ q_{b_{1}}(j)$ for all time intervals $j \in \mathcal{J}$ can be solved by: 
\begin{equation}\label{CP:lemma_connection}
\begin{array}{lll}
\underset{q_{b_{1}}}{\rm Maximize} & f(q_{b_{1}}) \\
{\rm s.t.}  &
\lbrace q_{b_{1}}(j) \mid \forall j \in \mathcal{J}\rbrace \in \mathcal{F}_{1, s}\quad \eqref{e:feasible_region_1},\\
&\lbrace q_{a_{2}}(j) \mid \forall j\in \mathcal{J} \rbrace  \in \mathcal{F}_{2,r}\quad \eqref{e:feasible_region_2},\\
&q_{b_{1}}(j) = q_{a_{2}}(j), \;\forall j \in \mathcal{J},
\end{array}
\end{equation}
if $f(q_{b_{1}})$ satisfies: $\forall j \in \mathcal{J}$,
\begin{equation}\label{e:lemma_connection_condition}
\begin{array}{l}
\frac{\partial f}{\partial q_{b_{1}}(1)\Delta t_{1}}  > \frac{\partial f}{\partial q_{b_{1}}(2)\Delta t_{2}} > \cdots > \frac{\partial f}{\partial q_{b_{1}}(j_{max})\Delta t_{j_{max}}} >0.
\end{array}
\end{equation}
\end{lemma}
The condition~\eqref{e:lemma_connection_condition} assigns a higher weight to the internal boundary flow at an earlier interval, hence satisfying the throughput maximization rule for all intervals. The proof of Lemma~\ref{lemma:connection} can be derived using the same technique in the proof for Proposition~\ref{prop:multiple_step} and is not detailed here.
}

Compared to a sequential scheme, such as the sequential convex program scheme or the CTM~\cite{daganzo1995net}, the single convex program framework allows a natural extension to optimal traffic control which is demonstrated in the Section~\ref{sec:on_ramp_control}.

\section{Application: On-ramp metering control for work zones} \label{sec:on_ramp_control}
This section demonstrates how the convex optimization scheme for computing the internal boundary flows can be reformulated as a control framework. An example for on-ramp metering control in work zones is provided. The proposed on-ramp metering controller uses historical data and the real-time measurement data at the entrance and exit of each road section to predict the traffic states in the work zone, and then avoids congestion upstream of the work zone by directly penalizing the congested traffic states.

\subsection{Optimal on-ramp metering control framework}
The convex optimization framework in the previous section computes the internal boundary flows on a network using the HJ PDE link model and proposed junction solver. Particularly, the junction models are encoded in the objective function in the convex program in Definition~\ref{prop:multiple_step_condition}. Without the conditions on the objective function in Definition~\ref{prop:multiple_step_condition}, the junction dynamics is no longer modeled and the internal boundary flow solution can be any value that is optimal for a given arbitrary objective function. In this sense, the convex program is an optimal controller assuming all directions of traffic are signalized at the junction. We refer to our earlier work~\cite{yanningControl} for a detailed discussion. 

The optimal controller for an on-ramp meter, however, is more complicated since both uncontrolled freeway flows and controlled on-ramp flows appear in the convex program. \edit{
The convex program CP~\eqref{CP:multiple_steps_merge} is used for solving the joint PDEs in the unsignalized merge, hence needed to be modified such that one upstream link (i.e., the on-ramp) allows control input. Meanwhile, the two freeway links preserve the unique internal boundary flow solution, which is similar to the problem of solving the joint PDEs at a connection junction in Lemma~\ref{lemma:connection}. Therefore, in the formulation of the optimal controller, the convex program for the connection~\eqref{CP:lemma_connection} is first used to guarantee the unique solution to the uncontrolled boundary flows. The remainder of this section shows how to include additional constraints and objectives into the convex program such that the on-ramp flows are controlled optimally to minimize the congestion. }

\subsection{Penalty on the congested states}
In the convex optimization framework, the congested states on links can be directly penalized via sampling the traffic condition at a set of points of interest $\mathcal{P}$ which we refer to as \emph{congestion sampling points}. For example, the set of congestion points is defined at discrete time points at a fixed location $x_{q}$, $i.e., \mathcal{P}:=\left\lbrace (t_{k}, x_{q}) \mid k \in \{1, 2, \cdots \} \right\rbrace$. If the congestion does not extend past the fixed location $x_{q}$, the traffic state is considered to be lightly congested and no penalty will be computed. Otherwise, the objective function penalizes the congested states depending on its severity. This subsection shows how this penalty mechanism is incorporated in the convex program by including additional constraints and variables. We briefly summarize the main ideas on the formulation of the constraints relying on the property of the partial solutions, and we refer to \cite{mazare2011analytical} for detailed interpretation of the property of partial solutions. 

By Proposition~\ref{prop:inf_morphism}, the Moskowitz solution at each congestion sampling point $(t_{k}, x_{q}) \in \mathcal{P}$ is computed as the minimum of the partial solutions computed from the affine initial conditions, the upstream boundary condition, or the downstream boundary condition. The affine initial conditions can be further grouped into two categories by whether the initial condition interval is in free flow or congested states, i.e., $ \rho_{l,0}(i) \leq \rho_{l}^{\kappa}$ or $\rho_{l,0}(i)>\rho_{l}^{\kappa}$, noting the density is related to the initial condition~\eqref{e:hj_icbc} by $\rho_{l,0}(i) = \frac{c_{l,0}^{i}(x_{i})-c_{l,0}^{i}(x_{i-1})}{\Delta x_{i}}$. Denote the sets of affine initial conditions that are in free flow and congested states respectively as $\mathcal{C}_{ff} = \{ c_{l,0}^{i}(x)\mid \rho_{l,0}(i) \leq \rho_{l}^{\kappa} \}$ and $\mathcal{C}_{cs} = \{ c_{l,0}^{i}(x) \mid \rho_{l,0}(i) > \rho_{l}^{\kappa} \}$ and the sets of affine upstream and downstream boundary conditions as $\mathcal{C}_{us} = \{ c_{a_{l}}^{j}(t), \forall j \in \mathcal{J} \}$,  and $\mathcal{C}_{ds} = \{ c_{b_{l}}^{j}(t), \forall j \in \mathcal{J} \} $. 

The partial solutions associated with $\mathcal{C}_{ff}$ and $\mathcal{C}_{us}$ imply free flow traffic conditions, and the partial solutions of $\mathcal{C}_{cs}$ and $\mathcal{C}_{ds}$ indicate congested traffic conditions in their respective characteristic solution domain. A congestion sampling point $(t_{k}, x_{q}) \in \mathcal{P}$ is in a free flow condition if and only if 
\begin{equation}\label{e:Mff}
\mathbf{M}_{l}(t_{k}, x_{q}) = \mathbf{M}_{ff}(t_{k}, x_{q}) := \min\left(\mathbf{M}_{l}^{c}(t_{k}, x_{q}), \forall c \in \mathcal{C}_{ff} \cup \mathcal{C}_{us} \right).
\end{equation}
In other words, the solution at $(t_{k}, x_{q})$ is in free flow if it is defined by the upstream condition or the free flow initial condition. Similarly the congestion sampling point is in a congested state if and only if
\begin{equation}\label{e:Mcs}
\mathbf{M}_{l}(t_{k}, x_{q}) = \mathbf{M}_{cs}(t_{k}, x_{q}):= \min\left(\mathbf{M}_{l}^{c}(t_{k}, x_{q}), \forall c \in \mathcal{C}_{cs} \cup \mathcal{C}_{ds} \right).
\end{equation} 

In addition, the partial solutions have the following properties: \textit{(i)} if a point $(t,x)$ is located in the characteristic domain of an affine initial or boundary condition, then the partial solution associated with the affine initial or boundary condition is equal to the solution $\mathbf{M}_{l}(t,x)$~\cite{mazare2011analytical}; \textit{(ii)} The characteristic solution domains for $c \in \mathcal{C}_{ff}\cup \mathcal{C}_{us}$ do not overlap, see~\eqref{e:LaxHopf}. Utilizing above properties, the nonlinear operator $\min$ in~\eqref{e:Mff} can be removed by pre-computing to which characteristic domain the point $(t_{k}, x_{q})$ belongs. Similarly, the $\min$ operator in~\eqref{e:Mcs} can be removed by applying the same technique.

For each congestion sampling point $(t_{k}, x_{q})$, the following constraints are added to the convex program~(\ref{CP:multiple_steps_merge}): 
\begin{equation}\label{e:congestion_constraints}
\begin{array}{l}
\forall (t_{k}, x_{q}) \in \mathcal{P},\\
\left\lbrace
\begin{array}{l}
\mathbf{M}_{ff}(t_{k}, x_{q}) \leq \mathbf{M}_{cs}(t_{k}, x_{q}) + p_{k},\\
p_{k} \geq 0,
\end{array}\right.
\end{array}
\end{equation} 
where $p_{k}$ is a variable that denotes the penalty associated with the congested states at each point $(t_{k}, x_{q})$. These constraints guarantee $p_{k} > 0$ when the point $(t_{k}, x_{q})$ is congested, i.e., $\mathbf{M}_{ff}(t_{k}, x_{q}) > \mathbf{M}_{cs}(t_{k}, x_{q})$, and $p_{k} = 0$ otherwise. The penalty variables $p_{k}$ for all congestion sampling points in $\mathcal{P}$ are regarded as decision variables and minimized in the objective function.

\subsection{Simulation configuration}
This subsection briefly describes the experiment setup and the next subsection constructs an optimal on-ramp metering controller using the convex program~\eqref{CP:lemma_connection} and the additional constraints~\eqref{e:congestion_constraints}.

\begin{figure}[htbp]
\begin{center}
\includegraphics[width=0.8\textwidth]{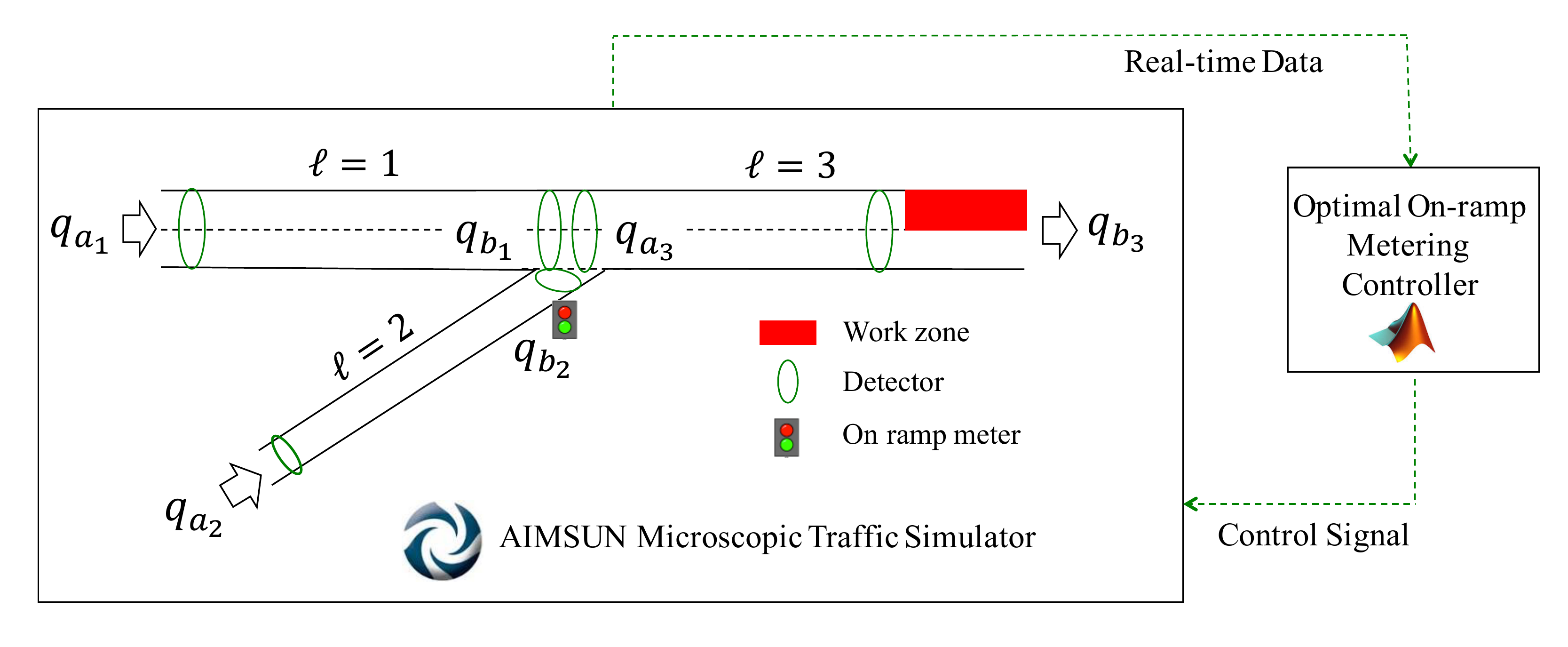}    
\caption{{On-ramp metering control for a work zone}: AIMSUN simulates a microscopic traffic environment, collects and feeds data to the optimal controller. The optimal controller implemented in MATLAB computes the optimal on-ramp signals and applies to AIMSUN.}  
\label{fig:aimsun_sim}                              
\end{center}                                 
\end{figure}

The experiment setup for validating the optimal on-ramp metering controller is shown in Fig.~\ref{fig:aimsun_sim}. A microscopic traffic simulation software, AIMSUN, is used to simulate a traffic environment and collect aggregated traffic data. \edit{The microscopic traffic simulator simulates the behavior of individual vehicles as a proxy for a real freeway network composed of human drivers, and is commonly used to validate traffic controllers based on macroscopic models~\cite{sun2005localized}.} The simulated data is then streamed to the optimal on-ramp metering controller implemented in MATLAB, which computes the optimal control signals based on the traffic dynamics modeled by the HJ PDE. Finally, the optimal on-ramp meter control is applied in the AIMSUN environment to simulate the evolution of traffic.

A six km stretch of freeway and an on-ramp is modeled in AIMSUN as in Fig.~\ref{fig:aimsun_sim}: a two-lane freeway (link~1) merges with a single-lane on-ramp (link~2) and connects to a two-lane freeway (link~3). A downstream work zone creates a single-lane bottleneck which induces congestion if the traffic is not controlled. Loop detectors are assumed to be installed at the entrance and exit of each link with a detection cycle set as 30 seconds. An on-ramp meter is installed at the exit of the on-ramp. Due to the limited work zone capacity, severe congestion will be generated upstream of the work zone if the on-ramp is not controlled. In comparison, the proposed controller can limit the level of congestion by regulating on-ramp inflows.

\begin{figure}[H]
\begin{center}
\includegraphics[width=0.8\textwidth]{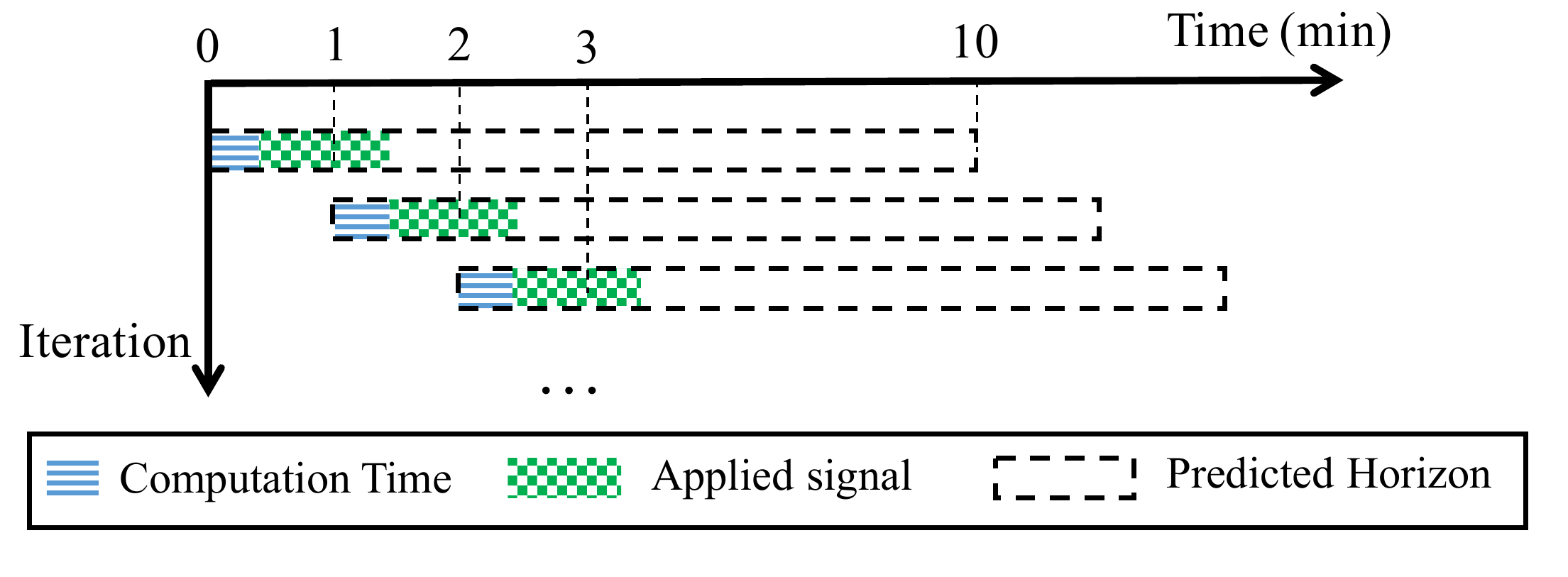}    
\caption{{MPC scheme for the on-ramp metering control}: Once new data is received, the optimal controller initializes a 10-minute predicted horizon for computing the optimal meter signals. New data is received each minute, hence only the first minute of the control signal during the 10-min horizon is applied to the on-ramp meter.}  
\label{fig:MPC}                              
\end{center}                                 
\end{figure}

The total simulated time horizon in this example is one hour. The controller is embedded in a \textit{model predictive control} (MPC) scheme~\cite{camacho2012model}~\cite{garcia1989model}~\cite{maciejowski2002predictive} which updates the control signals based on real-time measurements. The MPC scheme is illustrated in Fig.~\ref{fig:MPC}: \textit{(i)} the optimal controller predicts the traffic states over the next $10$-minute time horizon and computes the optimal ramp meter signal; \textit{(ii)} AIMSUN applies only the first minute of the control signal to the on-ramp meter, simulates the evolution of the traffic, and feeds back the aggregated traffic data to the controller; \textit{(iii)} the controller re-optimizes the ramp meter signal over the next $10$-minute time horizon using the new traffic measurement data from AIMSUN; \textit{(iv)} repeat steps \textit{(ii)} and \textit{(iii)} to adjust the optimal control signal to the realtime traffic measurement data.

In this example, the historical data is used for computing the optimal traffic signals over each $10$-minute time horizon. The error of the measurement data is not modeled in this example and we refer to our earlier work~\cite{li2014efficient} on robust optimal control with incorporates measurement uncertainty.

\subsection{Formulation of the Optimal Controller}
This subsection summarizes the constraints, designs the objective function, and formulates an optimal on-ramp metering controller. 

\subsubsection{Decision variables}
During each $10$-minute time horizon, the on-ramp boundary flows $q_{b_{2}}(j), \forall j\in \mathcal{J}$ are controlled and the freeway downstream flows $q_{b_{1}}(j), q_{b_{3}}(j), \forall j \in \mathcal{J}$ are computed using the junction model. Therefore, they are regarded as the decision variables in the convex program. 

The inflow to the downstream freeway link $l=3$ satisfies $q_{a_{3}}(j) = q_{b_{1}}(j) + q_{b_{2}}(j), \forall j \in \mathcal{J}$ and does not need to be explicitly included as a decision variable. The inflows to the upstream freeway and the on-ramp $q_{a_{1}}(j)$ and $q_{a_{2}}(j), \forall j\in \mathcal{J}$ are assumed to be known from the historical data.

In addition, the penalty $p_{k}$ associated with each congestion sampling point $(t_{k}, x_{q}) $ $ \in \mathcal{P}$ is also used as the decision variable to penalize congestion in the workzone.

In summary, the decision variable of the convex program is defined as follows:
\[\mathcal{X} = \left\lbrace q_{b_{1}}(j), q_{b_{2}}(j), q_{b_{3}}(j), p_{k} \mid \forall j \in \mathcal{J}, (t_{k}, x_{q})\in \mathcal{P} \right\rbrace . 
\]

\subsubsection{Linear constraints}
The constraints in the convex program for the optimal controller consist of the following linear inequalities and equalities.
\begin{itemize}
\item The boundary flows on each link $l$ subject to the feasible constraints $\mathcal{F}_{l, s}$ and $\mathcal{F}_{l, r}$ defined in equation~\eqref{e:feasible_region_1}~\eqref{e:feasible_region_2}.
\item The internal boundary flows at the on-ramp junction satisfy mass conservation for all time steps.
\item The congestion sampling points are selected as $\mathcal{P}= \lbrace \left( t_{k}, x_{q} \right)  \mid t_{k} =  30k\, s, \forall k \in \{0,1,2,\ldots\}$, $x_{q} = 50 \, m \rbrace$. The penalty variable associated with each congestion sampling point subjects to the constraints defined in equation~\eqref{e:congestion_constraints}.
\end{itemize}

\subsubsection{Objective function} One of the main safety concerns in work zones is the high-speed rear-end crashes which can be caused by the congestion upstream of the work zone. Therefore, the primary objective used in this example is to improve the safety for traveling through the work zone by alleviating the congestion on the freeway at the upstream of the work zone. Meanwhile, a secondary objective is to minimize the total travel time by sending on-ramp flow to the freeway as much as possible without causing congestion. The objective function $f$ is defined as a linear combination of several objective components. 
\begin{itemize}
\item The first component of the objective is to alleviate the congestion upstream of the work zone by directly penalizing the congested states at the congestion sampling points:
\[
\underset{p_{k}}{{\rm Maximize}} \qquad  - w_{0}\sum_{(t_{k}, x_{q}) \in \mathcal{P}}p_{k}, 
\]
where $w_{0}$ is a weight parameter which can be adjusted. 

\item The second component of the objective is to maximize the on-ramp flow which is metered by the controller:
\[
\underset{q_{b_{2}}(j)}{{\rm Maximize}}  \qquad  \sum_{j=1}^{j_{max}} w_{1}(j) q_{b_{2}}(j).
\]
Note if the weights $w_{1}(j)$ are the same for all time intervals $j \in \mathcal{J}$, the optimal controller may hold on-ramp flows to later time intervals which increases the waiting time of vehicles on the on-ramp. Therefore, we assign higher weights to on-ramp flows at earlier time intervals. Specifically, the weights $w_{1}(j)$ satisfy:
\begin{equation}\label{e:onramp_weight_condition}
w_{1}(j) > \frac{\Delta t_{j}}{\Delta t_{j+1}}w_{1}(j+1), \forall j \in \mathcal{J}\setminus \{j_{max}\}.
\end{equation}

\item In this example, the downstream boundary flows on the two freeway sections are not controlled. To obtain the unique solution at those two boundaries, the objective function $f$ must satisfy the conditions~\eqref{CP:lemma_connection}:
\begin{equation}\label{e:condition_for_connections}
\left\lbrace
\begin{array}{l}
\frac{\partial f}{\partial q_{b_{1}}(j)} > \frac{\Delta t_{j}}{\Delta t_{j+1}} \frac{\partial f}{\partial q_{b_{1}}(j+1)},\\
\frac{\partial f}{\partial q_{b_{3}}(j)} > \frac{\Delta t_{j}}{\Delta t_{j+1}} \frac{\partial f}{\partial q_{b_{3}}(j+1)}.
\end{array}
\right.
\end{equation}
The following objective component is added to the objective function to guarantee the unique solution:
\[
\underset{q_{b_{1}}(j), q_{b_{3}}(j)}{{\rm Maximize}} \qquad \sum_{j = 1}^{j_{max}} w_{2}(j)q_{b_{1}}(j)  + \sum_{j = 1}^{j_{max}} w_{3}(j)q_{b_{3}}(j).
\]
It should be noted that $p_{k}$ relates to $q_{b_{1}}(j)$ in~\eqref{e:congestion_constraints} and minimizing $p_{k}$ implicitly minimizes $q_{b_{1}}(j)$. The selection of $w_{2}(j)$ using conditions~\eqref{e:condition_for_connections} should subtract the implicit weight on $q_{b_{1}}(j)$ induced by penalizing the congested states. In addition, to prevent the on-ramp flows from blocking the upstream freeway flows, the objective function should satisfy:
\begin{equation}\label{e:non_competing_condition}
\frac{\partial f}{\partial q_{b_{1}}(j)} > \frac{\partial f}{\partial q_{b_{2}}(j)}.
\end{equation}
\end{itemize}

In summary, the objective function $f$ is to maximize:
\[
f = - w_{0}\sum_{(t_{k}, x_{q}) \in \mathcal{P}}p_{k} + \sum_{j=1}^{j_{max}} w_{1}(j) q_{b_{2}}(j) + \sum_{j = 1}^{j_{max}} w_{2}(j)q_{b_{1}}(j)  + \sum_{j = 1}^{j_{max}} w_{3}(j)q_{b_{3}}(j).
\]
Specifically in this simulation, we select $w_{0} = 1$, $w_{1}(j_{max}) = \Delta t_{j_{max}}$, and other weights according to conditions~\eqref{e:onramp_weight_condition},~\eqref{e:condition_for_connections},~\eqref{e:non_competing_condition}. The weights $w_0$ and $w_1$ can be adjusted to balance the congestion on the highway and the queue on the on-ramp.

\subsection{Simulation results}
\begin{figure}[h]
 	\begin{subfigure}{\textwidth}
        \centering
        \includegraphics[width=0.65\textwidth]		{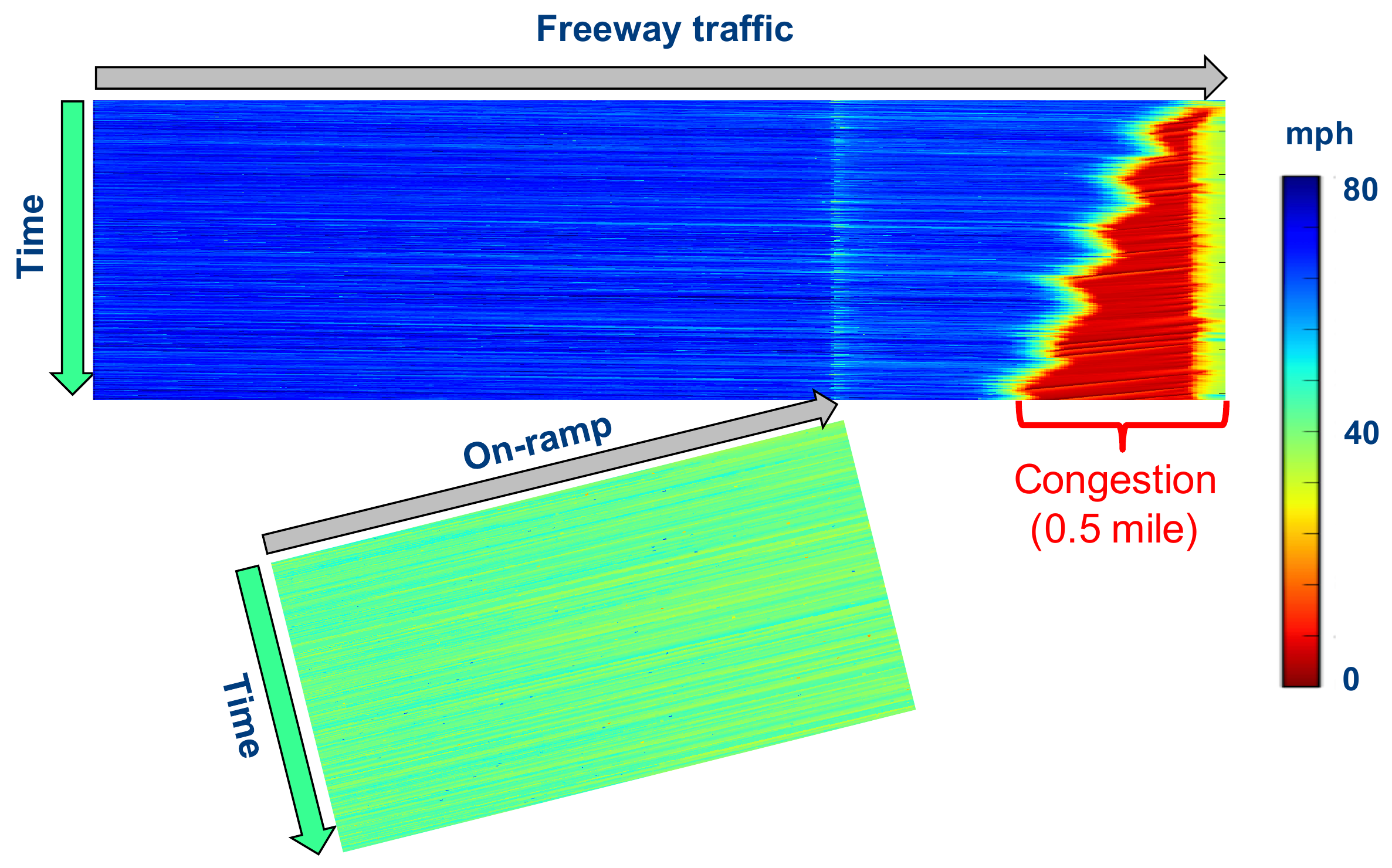}
        \caption{Uncontrolled on-ramp. On-ramp flows caused severe congestion on the downstream freeway link before the work zone bottleneck.}
        \label{fig:onramp_uncontrolled}
    \end{subfigure}
    
    \begin{subfigure}{\textwidth}
        \centering
        \includegraphics[width=0.65\textwidth]		{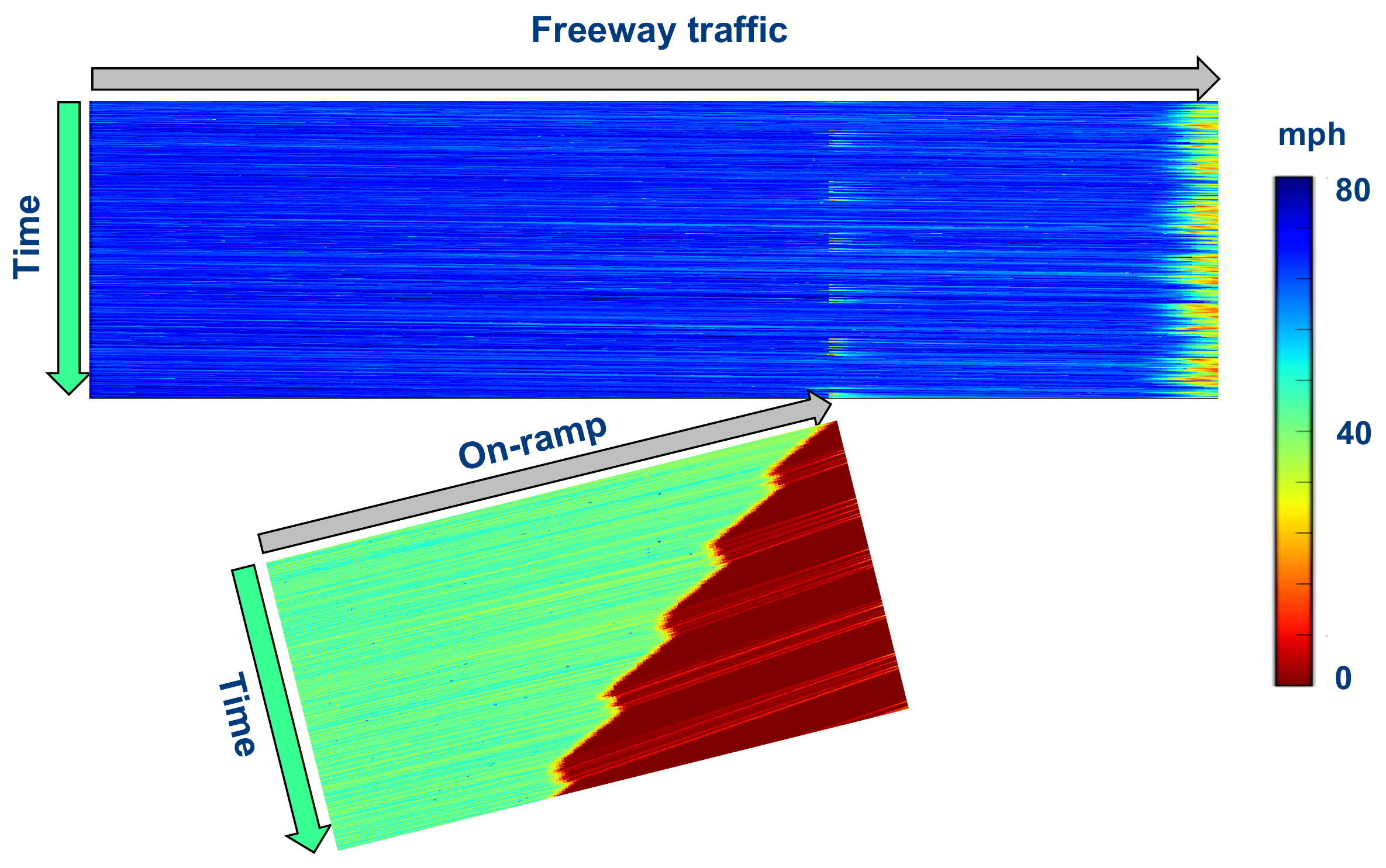}
        \caption{Optimal on-ramp control. On-ramp flows were restricted to prevent the formation of congestion on the downstream freeway link. Meanwhile, on-ramp flows were admitted to the freeway when there was space on the downstream freeway link. Overall, no severe congestion was generated on the downstream freeway link except for slightly slower traffic due to the reduced speed limit and merging activities upstream of the work zone.}
        \label{fig:onramp_controlled}
    \end{subfigure}

\caption{True speed states on the freeway and the on-ramp. Grey arrows denote the freeway and on-ramp links. Green arrow denote the time (1 hr). (a) Un-controlled on-ramp. (b) Optimally controlled on-ramp.}  
\label{fig:aimsun_sim_result}                              
\end{figure}

A one-hour time horizon was simulated in AIMSUN for the optimal on-ramp metering control. A scenario with uncontrolled on-ramp meter was also simulated for comparison. The simulation result is shown in Fig.~\ref{fig:aimsun_sim_result}, where Fig.~\ref{fig:onramp_uncontrolled} plots the traffic states for the uncontrolled scenario and Fig.~\ref{fig:onramp_controlled} for the optimal on-ramp metering control scenario. 

As shown in Fig.~\ref{fig:onramp_uncontrolled}, the on-ramp flow is not controlled and all on-ramp traffic merges to the downstream freeway. On the downstream freeway, the work zone reduced the road capacity and caused severe congestion which could cause safety issues. In comparison in Fig.~\ref{fig:onramp_controlled}, the optimal on-ramp metering controller regulated the on-ramp traffic to the downstream freeway such that no severe congestion formed. The additional delay time of the vehicles waiting on the on-ramp is compensated by the shorter travel time on the uncongested downstream link. 

In summary, this section demonstrated the feasibility of reformulating the convex optimization scheme to optimal traffic control applications. The general idea is to relax the junction models encoded by the conditions on the objective function in Proposition~\ref{prop:multiple_step_condition}. A variety of objectives, such as maximizing the boundary flow, or penalizing the congested states can be directly formulated in the objective function. 

\section{Conclusion}
This article proposed a numerical scheme which can compute the traffic evolution modeled by HJ PDEs on a network using a convex optimization program, which could also be applied for optimal control. 

The proposed framework relies on a semi-explicit single link HJ PDE solver, and does not require discretization of the time-space domain. In addition, it computes the internal boundary flows at a merge, or diverge, \edit{or connection} over the entire time horizon using a single convex program. The convex optimization scheme provides a natural framework for optimal traffic control applications which is demonstrated in a work zone on-ramp metering control example.

\newpage
\section{Appendix}
\subsection{Discussion of the unique solution at the diverge}
\label{appendix:diverge}
\edit{
The unique solution during a single time interval $j$ at a diverge can be computed by a diverge junction solver which can be constructed following the construction of CP~\eqref{CP:single_step_merge} for the merge. Since the diverge junction solver is very similar to the merge junction solver except for notation changes, we only discuss the resulting unique solutions without articulating the mathematical details of the diverge junction solver.}

\begin{figure}[htpb]
    \begin{subfigure}{0.32\textwidth}
        \centering
        \includegraphics[width=\textwidth]{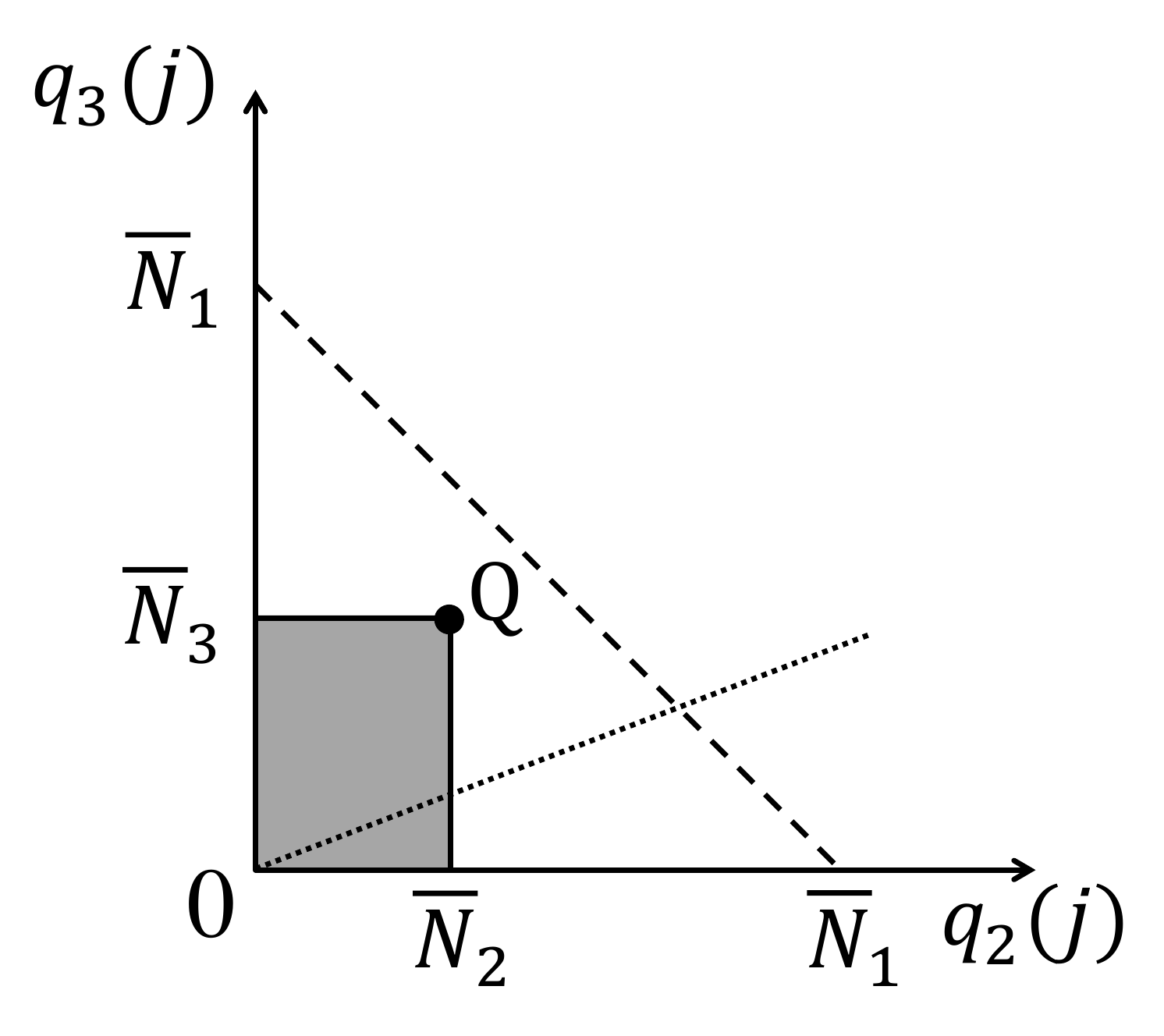}
        \caption{\textbf{Diverge Case One}}
        \label{fig:diverge_1}
    \end{subfigure}
    \begin{subfigure}{0.32\textwidth}
        \centering
        \includegraphics[width=\textwidth]{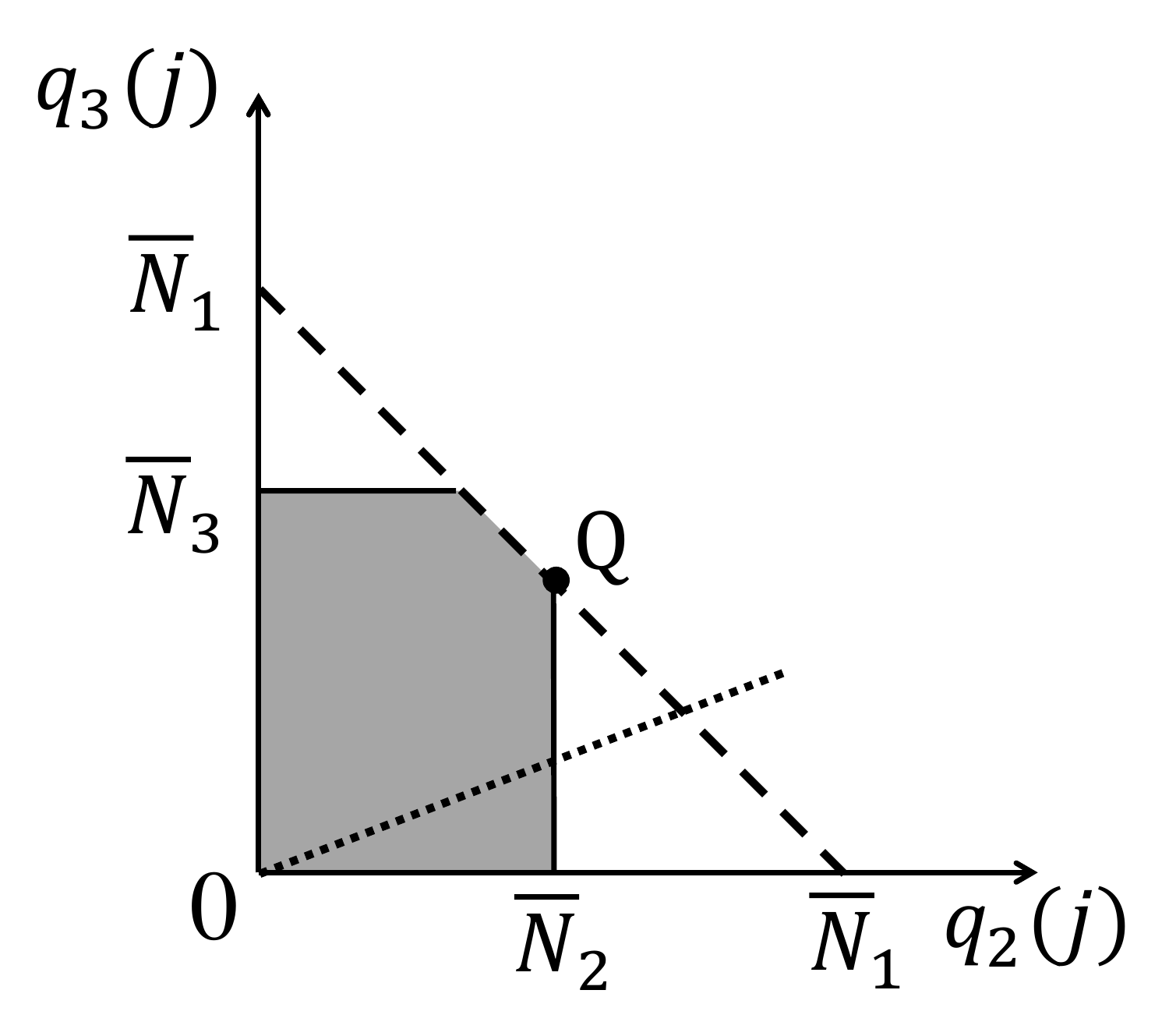}
        \caption{\textbf{Diverge Case Two}}
        \label{fig:diverge_2}
    \end{subfigure} 
    \begin{subfigure}{0.32\textwidth}
        \centering
        \includegraphics[width=\textwidth]{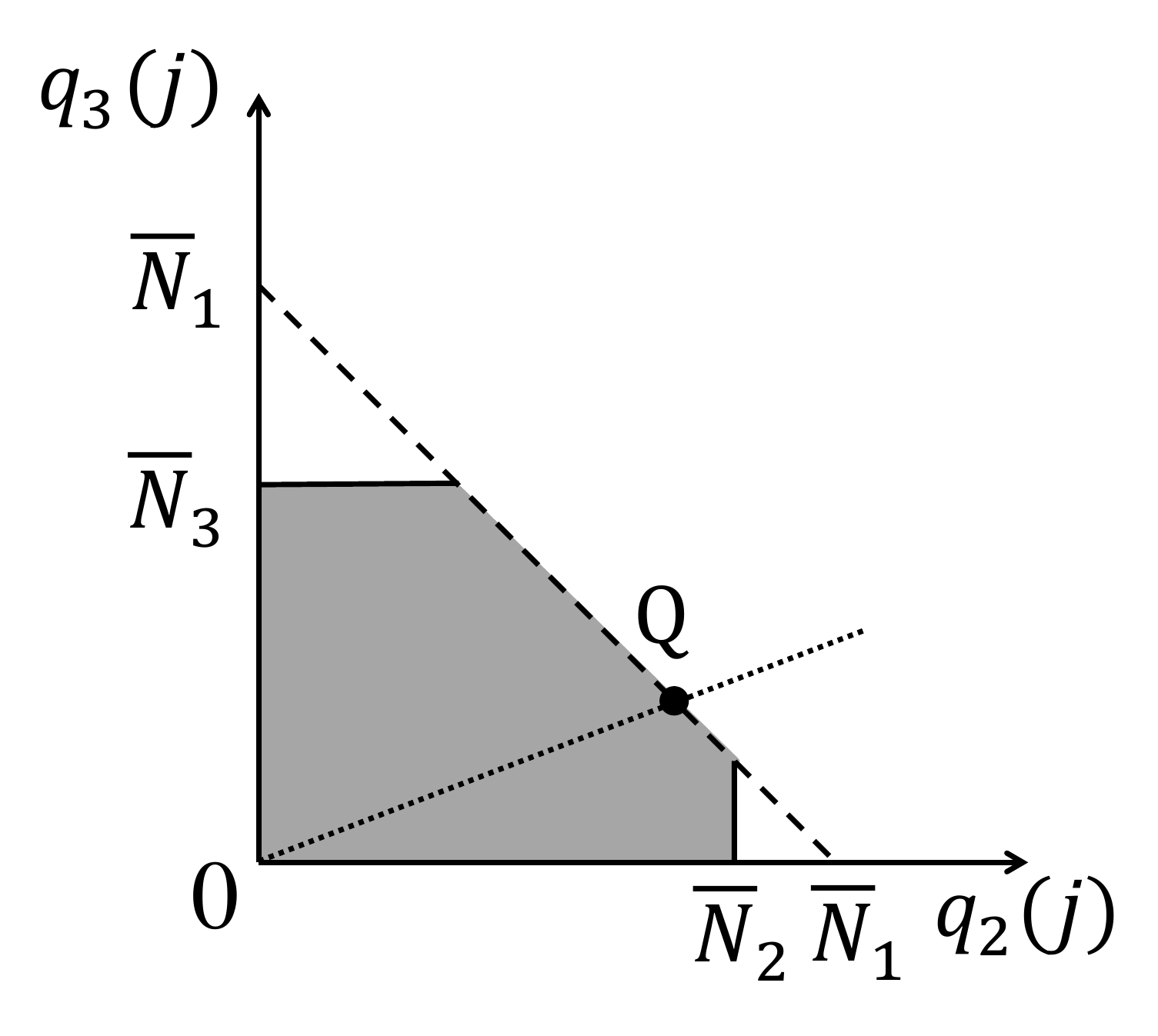}
        \caption{\textbf{Diverge Case Three}}
        \label{fig:diverge_3}
    \end{subfigure} 
    \caption{{Three scenarios at a diverge.} Link 1 diverges to link 2 and 3. The two solid lines represent the maximum receiving flow on links 2 and 3. The dashed line denotes the maximum sending flow on link 1. The dotted line denotes the prescribed distribution ratio. The shaded area is the feasible set of boundary flows~\eqref{e:single_step_feasible_region_1}~\eqref{e:single_step_feasible_region_2}. The unique solution computed by the diverge junction solver is depicted by Q.}
    \label{fig:Diverge_scenarios}
\end{figure}

The unique solution of the diverge junction model is illustrated in Fig.~\ref{fig:Diverge_scenarios}. There are in total three scenarios depending on the feasible sets on links~\eqref{e:single_step_feasible_region_1}~\eqref{e:single_step_feasible_region_2}, namely, \textit{(i)} when the maximum sending flow from link 1 exceeds the total receiving flow on links 2 and 3 in Fig.~\ref{fig:diverge_1}; \textit{(ii)} when the maximum sending flow is smaller than the total receiving flow on links 2 and 3 but the prescribed distribution ratio can not be followed exactly in Fig.~\ref{fig:diverge_2}; \textit{(iii)} when the maximum sending flow is smaller than the total receiving flow on links 2 and 3 and the prescribed distribution ratio can be followed exactly in Fig.~\ref{fig:diverge_3}. The feasible set of the convex program is denoted by the shaded area, where the upper bounds of the feasible internal boundary flows for three links are respectively  $\bar{N}_{1} = \max \left\lbrace { q_{1}(j) \mid  q_{1}(j)\in \mathcal{F}_{1, s}^{j}} \right\rbrace, \, \bar{N}_{2} = \max \left\lbrace q_{2}(j) \mid q_{2}(j)\in \mathcal{F}_{2, r}^{j}\right\rbrace$, and $\bar{N}_{3} = \max \left\lbrace q_{3}(j) \mid q_{3}(j)\in \mathcal{F}_{3, r}^{j} \right\rbrace$. The solid lines denote the maximum receiving flows on links 2 and 3. The dashed line denotes the maximum sending flow from link 1. The dotted line denotes the prescribed distribution of the boundary flows, i.e., $q_{2}(j) = \frac{ q_{1}(j) }{1+D},\; q_{3}(j) = \frac{Dq_{1}(j) }{1+D}$. The unique solution computed by a diverge junction solver is marked at point $Q$. 

In scenario Fig.~\ref{fig:diverge_1}, the maximum sending flow from link 1 exceeds the the total flow that can be received by links 2 and 3 combined. Hence, the single point that maximizes the throughput saturates the maximum receiving flow on links 2 and 3, and is the optimal solution to the diverge junction solver. 
It should be noted that the optimal solution $Q$ may not fall on the dotted distribution line, meaning distribution rule~\ref{r:3} is relaxed and a portion of vehicles originally headed to link 2 are rerouted to link 3. The benefit of such a model is that it prevents a blocked exit ramp from completely blocking all flows across the junction when applied to multilane freeways.

If the downstream links combined can receive more flow than the upstream link 1 can send, the maximum sending flow becomes an active constraint as shown in Fig.~\ref{fig:diverge_2}. There are an infinite number of flow maximizing solutions on the dashed line within the feasible set, and none of the solutions satisfy the prescribed distribution ratio  exactly. In this case, the sending flow from link 1 saturates the link 2 first, and then the vehicles that can not be admitted to link 2 will reroute to link 3. Consequently, the optimal solution to the diverge junction solver is the solution $Q$ that is closest to the dotted line among the solutions on the dashed line within the feasible set. 

In the last scenario, the total downstream links combined is higher than the maximum sending flow from link 1, and the sending flow can be distributed exactly following the distribution ratio as in Fig.~\ref{fig:diverge_3}. In this case, there is no conflicts between maximizing the throughput~\ref{r:2} and following the flow distribution~\ref{r:3}. Therefore, the solution at $Q$ that satisfies both~\ref{r:2} and~\ref{r:3} is the optimal solution to the diverge junction solver.

\newpage
\subsection{Proof of Proposition~\ref{def:junction_solution_merge}}\label{appendix:single}
The merge junction solver CP~\eqref{CP:single_step_merge} computes the unique internal boundary flow solution $q^{\ast}(j) = \left( q^{\ast}_{b_{1}}(j), q^{\ast}_{b_{2}}(j), q^{\ast}_{a_{3}}(j) \right)$ at interval $j$, where $q^{\ast}(j)$ satisfies \ref{r:1_merge},~\ref{r:2_merge},~\ref{r:3_merge}:
\begin{itemize}
\item[] \ref{c:1_merge} The internal boundary flows satisfy mass conservation~\ref{r:1_merge},
$
q^{\ast}_{a_{3}}(j) = q^{\ast}_{b_{1}}(j) + q^{\ast}_{b_{2}}(j).
$
\item[] \ref{c:2_merge} The throughput flow at the junction is maximized subject to the feasible sets on connecting links~\ref{r:2_merge}, i.e., $q^{\ast}(j) \in \mathcal{Q}_{j} := \underset{q(j) \in \mathcal{F}^{j}_{1,s} \times \mathcal{F}^{j}_{2,  s} \times \mathcal{F}^{j}_{3, r}}{ \operatorname{argmax} } q_{a_{3}}(j)$.
\item[]  \ref{c:3_merge} The deviation from the distribution equation is minimized~\ref{r:3}, i.e., $q^{\ast}(j) = \underset{q(j)\in \mathcal{Q}_{j}}{\operatorname{argmin} }$  $\| q_{b_{2}}(j) - Pq_{b_{1}}(j) \|_{1}$.
\end{itemize}


\begin{proof}
At a merge junction, there are in total three scenarios, \textit{(i)} the downstream link has sufficient capacity; \textit{(ii)} the downstream link has insufficient capacity, however there exists no solution that maximizes the throughput and satisfies the prescribed priority ratio at the same time; \textit{(iii)} the downstream has insufficient capacity, and there exists a solution that maximizes the throughput and satisfies the prescribed priority ratio at the same time. 

Denote the upper bounds of the sending and receiving boundary flows on three links at the junction as {\small $\bar{N}_{1} = \max \left\lbrace { q_{1}(j) \mid  q_{1}(j)\in \mathcal{F}_{1, s}^{j}} \right\rbrace$}, {\small $\bar{N}_{2} = \max \left\lbrace q_{2}(j) \mid q_{2}(j)\in \mathcal{F}_{2, s}^{j}\right\rbrace$}, and {\small $\bar{N}_{3} = \max \left\lbrace q_{3}(j) \right.$ $\left.\mid q_{3}(j)\in \mathcal{F}_{3, r}^{j} \right\rbrace$}.

\edit{By construction of the junction solver, the constraint set of the convex program is non-empty, which guarantees the existence of a solution.}
Suppose the \edit{optimal} solution is $\left(q_{1}^{\ast}(j), q_{2}^{\ast}(j) \right)$. It suffices to prove the objective function value $f\left(\hat{q_{1}}(j), \hat{q_{2}}(j)\right)$ associated with any feasible alternative solution $\left(\hat{q_{1}}(j), \hat{q_{2}}(j)\right)$ is strictly smaller than $f\left(q_{1}^{\ast}(j), q_{2}^{\ast}(j) \right)$. \\
\begin{itemize}
\item[] \emph{Scenario 1 (Fig.~\ref{fig:merge_1})}: The downstream link has sufficient space for the upstream sending flows. Then, the unique solution is obtained by admitting all the vehicles from the upstream links, i.e., $q_{1}^{\ast}(j) = \bar{N}_{1}$, and $q_{2}^{\ast}(j) = \bar{N}_{2}$. \edit{Since the the convex objective function is strictly and monotonically increasing in $[0, \bar{N}_1]\times [0, \bar{N}_2]$ by condition~(\ref{eq:prop_single_step_merge_1}), the optimal solution computed by the convex program is $(\bar{N}_1, \bar{N}_2)$.}



\item[] \emph{Scenario 2 (Fig.~\ref{fig:merge_2})}: The downstream link has insufficient space for the sending flows from the upstream links. However, there exists no solution that maximizes the flow while satisfying the priority parameter. Therefore, the unique solution is obtained by selecting the point on the dashed line segment in the feasible set that is closest to the prescribed priority parameter (dotted line), i.e., $q_{1}^{\ast}(j) = \bar{N}_{1}, q_{2}^{\ast}(j) = \bar{N}_{3} - \bar{N}_{1}$. By applying the same technique in Scenario 1 using  condition~(\ref{eq:prop_single_step_merge_1}),  we can observe that: $\forall  (\hat{q_{1}}(j), \hat{q_{2}}(j))$ not on the dashed line segment in the feasible set, $\exists (q'_{1}(j), q'_{2}(j))$ on the dashed line segment in the feasible set, such that the following inequality holds:
\[
 f(q'_{1}(j) , q'_{2}(j)) > f(\hat{q_{1}}(j), \hat{q_{2}}(j)).
\]
Therefore, it suffices to only consider the feasible solutions on the dashed line segment in the feasible set. Define an alternative solution on the dashed line segment as $\left( q_{1}^{\ast}(j)+\delta_{1}\right.$,$\left. q_{2}^{\ast}(j)\right.$ $\left.+\delta_{2} \right)$, where $\delta_{1}+\delta_{2} = 0$. In addition, since the alternative solution is a feasible solution, i.e., $q_{1}^{\ast}(j)+\delta_{1} \leq \bar{N}_{1}$, hence $\delta_{1} \leq 0, $ The objective function value for the alternative solution is
{\small
\[
\begin{array}{ll}
f(q_{1}^{\ast}(j) + \delta_{1}, q_{2}^{\ast}(j)+ \delta_{2}) &= f(q_{1}^{\ast}(j) , q_{2}^{\ast}(j)) + \frac{\partial f}{\partial q_1(j)}\delta_{1} + \frac{\partial f}{\partial q_2(j)} \delta_{2},\\
& \leq 
f(q_{1}^{\ast}(j) , q_{2}^{\ast}(j)) + (\delta_{1}+ \delta_{2})\frac{\partial f}{\partial q_2(j)}, \qquad \text{by~(\ref{eq:prop_single_step_merge_2}) }, \\
& = f(q_{1}^{\ast}(j) , q_{2}^{\ast}(j)).
\end{array}
\]
}
The equality holds only when $\delta_{1} = \delta_{2} = 0$, which proves the uniqueness of the solution.\\

\item[] \emph{Scenario 3 (Fig.~\ref{fig:merge_3})}: In this scenario, the downstream link has sufficient space for the sending flows from the upstream links, and there exists a solution that maximizes the flow while satisfying the priority parameter. The unique solution is obtained by distributing the maximum sending flow from the upstream link using the distribution parameter $P$, i.e., $q_{1}^{\ast}(j) = \bar{N}_3/(1+P), q_{2}^{\ast}(j) = \bar{N}_{3}\cdot P/(1+P)$. Similarly as in Scenario 2, it suffices to only consider alternative solutions on the dashed line segment in the feasible set $\left( q_{1}^{\ast}(j)+\delta_{1} , q_{2}^{\ast}(j)+\delta_{2} \right)$, where by definition $\delta_{1} + \delta_{2} = 0$. 
\begin{itemize}
\item If $\delta_{1} \leq 0$, then $\delta_{2}  = -\delta_{1} \geq 0$, and $q_{2}^{\ast}(j) + \delta_{2} = Pq_{1}^{\ast}(j) + \delta_{2} \geq P(q_{1}^{\ast}(j) + \delta_{1})$. The corresponding objective function value is
{\footnotesize
\[
\begin{array}{ll}
f(q_{1}^{\ast}(j) + \delta_{1}, q_{2}^{\ast}(j) + \delta_{2}) &= f(q_{1}^{\ast}(j), q_{2}^{\ast}(j)) + \frac{\partial f}{\partial q_{1}(j)}\delta_{1} + \frac{\partial f}{\partial q_{2}(j)} \delta_{2},\\
& \leq f(q_{1}^{\ast}(j) , q_{2}^{\ast}(j)) + (\delta_{1}+ \delta_{2})\frac{\partial f}{\partial q_{1}(j)}, \qquad \text{by~(\ref{eq:prop_single_step_merge_2})},\\
& \leq f(q_{1}^{\ast}(j) , q_{2}^{\ast}(j)).
\end{array}
\]
}
The equality holds only when $\delta_{1} = \delta_{2} = 0$.
\item If $\delta_{1} > 0$, then $\delta_{2} = -\delta_{1} < 0$, and $q_{2}^{\ast}(j) + \delta_{2} = Pq_{1}^{\ast}(j) + \delta_{2} \leq P(q_{1}^{\ast}(j) + \delta_{1})$. The corresponding objective function value is
{\footnotesize
\[
\begin{array}{ll}
f(q_{1}^{\ast}(j) + \delta_{1}, q_{2}^{\ast}(j) + \delta_{2}) &= f(q_{1}^{\ast}(j) , q_{2}^{\ast}(j)) + \frac{\partial f}{\partial q_{1}(j)}\delta_{1} + \frac{\partial f}{\partial q_{2}(j)} \delta_{2},\\
& < f(q_{1}^{\ast}(j), q_{2}^{\ast}(j)) + (\delta_{1}+ \delta_{2})\frac{\partial f}{\partial q_{2}(j)}, \qquad \text{by~(\ref{eq:prop_single_step_merge_3}) },\\
& < f(q_{1}^{\ast}(j), q_{2}^{\ast}(j)).
\end{array}
\]
}
\end{itemize}
Therefore, the objective value for all feasible points other than the solution $(q_{1}^{\ast}(j), q_{2}^{\ast}(j))$ are strictly smaller, and therefore $(q_{1}^{\ast}(j), q_{2}^{\ast}(j))$ is unique. 
\end{itemize}
\end{proof}

\subsection{Proof of Proposition~\ref{prop:multiple_step}}\label{appendix:multiple}
The junction solver~\eqref{CP:multiple_steps_merge} gives the same unique solution $\left\lbrace q_{b_{1}}(j), q_{b_{2}}(j), q_{a_{3}}(j) \mid \forall j \in \mathcal{J} \right\rbrace$ obtained by sequentially solving CP~\eqref{CP:single_step_merge} at each time interval.

\begin{proof}
The proof relies on the equivalence of the KKT conditions between the sequence of convex program for each single time interval and the single convex program for the entire time horizon. The proof is presented in four steps. \textit{Step 1:} write the KKT conditions of convex programs for each single time interval. \textit{Step 2:} remove the terms on the objective function in the KKT conditions using the conditions~\eqref{eq:prop_single_step_merge}. \textit{Step 3:} write the KKT conditions of the single convex program over the entire time horizon. \textit{Step 4:} show the optimal solution and associated multipliers that satisfy the KKT conditions of the single convex program over the entire time horizon and the proposed conditions~\eqref{eq:prop_multiple_step_12}~\eqref{eq:prop_multiple_step_34}~\eqref{eq:prop_multiple_step_56} also satisfy the KKT conditions for each time interval, hence the solution is the unique solution at each time interval.\\

\textbf{\emph{Step 1: Write the KKT condtions for CP~(\ref{CP:single_step_merge})} at each time interval}.

For each interval $j \in \mathcal{J}$, assuming the unique solution until interval $j-1$ are known, we can rewrite CP~(\ref{CP:single_step_merge}) explicitly as follows:
{\footnotesize
\begin{equation}\label{CP:proofonestep}
\begin{array}{ll}
\underset{q_{1}(j), q_{2}(j)}{\rm Maximize}& \quad f(q_{1}(j), q_{2}(j)) \\
{\rm s.t.} &\quad \sum_{\tau=1}^{j-1} q_{1}^{\ast}(\tau)\Delta t_{\tau} + q_{1}(j)\Delta t_{j} \leq \bar{N}_{1}(j),\\
&\quad \sum_{\tau=1}^{j-1} q_{2}^{\ast}(\tau)\Delta t_{\tau} + q_{2}(j)\Delta t_{j} \leq \bar{N}_{2}(j),\\
&\quad \sum_{\tau=1}^{j-1} q_{3}^{\ast}(\tau)\Delta t_{\tau} + q_{3}(j)\Delta t_{j} \leq \bar{N}_{3}(j),\\
&\quad q_{1}(j) \leq q_{1}^{max}, \\
&\quad q_{2}(j) \leq q_{2}^{max},\\
&\quad q_{3}(j) \leq q_{3}^{max}, \\
&\quad  q_{1}(j) + q_{2}(j) = q_{3}(j).
\end{array}
\end{equation}
}

The KKT conditions for above CP can be written as follows. If $\left( q_{1}^{\ast}(j), q_{2}^{\ast}(j), q_{3}^{\ast}(j) \right)$ is the optimal solution of CP~(\ref{CP:proofonestep}), then there exist multipliers $\lambda_{i}, i \in \{1, \cdots, 6\}$, such that 
{\footnotesize
\begin{equation}\label{eq:singleKKT}
\left\{
\begin{array}{l}
\lambda_i \geq 0, \quad \forall i \in \{1,2\ldots 6\},\\
\\
\lambda_1 \cdot\left( \sum_{\tau=1}^{j}q_{1}^{\ast}(\tau)\Delta t_{\tau} - \bar{N}_{1}(j) \right) = 0,\\
\lambda_2 \cdot\left( \sum_{\tau=1}^{j}q_{2}^{\ast}(\tau)\Delta t_{\tau} - \bar{N}_{2}(j)\right) = 0, \\
\lambda_3 \cdot\left( \sum_{\tau=1}^{j}\left(q_{1}^{\ast}(\tau) + q_{2}^{\ast}(\tau) \right)\Delta t_{\tau} - \bar{N}_{3}(j) \right) = 0, \\
\lambda_{4} \cdot \left( q_{1}^{\ast}(j) - q_{1}^{max} \right) = 0, \\
\lambda_{5} \cdot \left( q_{2}^{\ast}(j) - q_{2}^{max} \right) = 0, \\
\lambda_{6} \cdot \left( q_{1}^{\ast}(j) + q_{2}^{\ast}(j) - q_{3}^{max} \right) = 0, \\
\\
-\frac{\partial f}{\partial q_{1}(j)\Delta t_{j}} + \lambda_1 + \lambda_3 + \lambda_4 + \lambda_6 = 0, \\
-\frac{\partial f}{\partial q_{2}(j)\Delta t_{j}} + \lambda_2 + \lambda_3 + \lambda_5 + \lambda_6 = 0. \\
\end{array} 
\right.
\end{equation}
}

Note that since there are only linear constraints in CP~(\ref{CP:proofonestep}), the KKT conditions are also sufficient conditions. Hence, given boundary flows $(q_{1}(j), q_{2}(j))$, if there exists a set of multipliers that satisfies the KKT conditions above for $(q_{1}(j), q_{2}(j))$, then $(q_{1}(j), q_{2}(j))$ is the optimal solution.\\

\textbf{\emph{Step 2: Combine the conditions~\eqref{eq:prop_single_step_merge} with the KKT conditions for CP~(\ref{CP:proofonestep})}}.

From \emph{Step 1}, the derivatives of the objective function ${\partial f}/{\partial q_{1}(j)}, {\partial f}/{\partial q_{2}(j)}$ are related to the set of multipliers by the stationarity condition in the KKT conditions. 

Combined with the conditions on $f$~\eqref{eq:prop_single_step_merge} from the junction solver~\eqref{CP:single_step_merge}, the above KKT conditions can be rewritten as follows. If $\left( q_{1}^{\ast}(j), q_{2}^{\ast}(j)\right)$ is the unique solution of CP~(\ref{CP:proofonestep}), then there exist $\lambda_{i}, i \in \{1, \cdots, 6\}$, such that:
{\footnotesize
\[
(\text{KKT}_{j}) \left\{
\begin{array}{ll}
\lambda_i \geq 0 \quad \forall i \in \{1,2\ldots 6\}, &\\
\\
\lambda_1 \cdot\left( \sum_{\tau=1}^{j}q_{1}^{\ast}(\tau)\Delta t_{\tau} - \bar{N}_{1}(j) \right) = 0,\\
\lambda_2 \cdot\left( \sum_{\tau=1}^{j}q_{2}^{\ast}(\tau)\Delta t_{\tau} - \bar{N}_{2}(j)\right) = 0, \\
\lambda_3 \cdot\left( \sum_{\tau=1}^{j}\left(q_{1}^{\ast}(\tau) + q_{2}^{\ast}(\tau) \right)\Delta t_{\tau} - \bar{N}_{3}(j) \right) = 0, \\
\lambda_{4} \cdot \left( q_{1}^{\ast}(j) - q_{1}^{max} \right) = 0, \\
\lambda_{5} \cdot \left( q_{2}^{\ast}(j) - q_{2}^{max} \right) = 0, \\
\lambda_{6} \cdot \left( q_{1}^{\ast}(j) + q_{2}^{\ast}(j)  - q_{3}^{max} \right) = 0, \\
\\
\lambda_1 + \lambda_3 + \lambda_4 + \lambda_6 > 0, \\
\lambda_2 + \lambda_3 + \lambda_5 + \lambda_6 > 0, \\
\lambda_1 + \lambda_3 + \lambda_4 + \lambda_6 > \lambda_2 + \lambda_3 + \lambda_5 + \lambda_6, & \text{ if } q_{2}(j) \geq Pq_{1}(j), \\
\lambda_1 + \lambda_3 + \lambda_4 + \lambda_6 < \lambda_2 + \lambda_3 + \lambda_5 + \lambda_6, & \text{ if } q_{2}(j) < Pq_{1}(j). \\
\end{array} 
\right.\\
\]
}

It should be noted that, compared with the original KKT conditions~\eqref{eq:singleKKT}, KKT$_j$ replaces the stationarity condition on the objective function $f$ by inequality constraints on the multipliers. \\

\textbf{\emph{Step 3: Write the KKT conditions for single CP~\eqref{CP:multiple_steps_merge} over the entire time horizon}}.

Similarly, CP~\eqref{CP:multiple_steps_merge} can be written explicitly in the following form:
{\footnotesize
\begin{equation}\label{CP:proofmultiplestep}
\begin{array}{ll}
\underset{q_{1}, q_{2}}{\rm Maximize}& \quad f(q_{1}, q_{2})  \\
{\rm s.t.} &\quad \forall j \in \mathcal{J},\\
& \quad \sum_{\tau=1}^{j} q_{1}(\tau)\Delta t_{\tau} \leq \bar{N}_{1}(j),\\
& \quad \sum_{\tau=1}^{j} q_{2}(\tau)\Delta t_{\tau} \leq \bar{N}_{2}(j),\\
& \quad \sum_{\tau=1}^{j} q_{3}(\tau)\Delta t_{\tau} \leq \bar{N}_{3}(j),\\
& \quad q_{1}(j) \leq q_{1}^{max}(j), \\ 
& \quad q_{2}(j) \leq q_{2}^{max}(j), \\ 
& \quad q_{3}(j) \leq q_{3}^{max}(j) , \\ 
&\quad  q_{1}(j) + q_{2}(j) = q_{3}(j).
\end{array}
\end{equation}
}

The associated KKT conditions can be stated as: if $\left\lbrace \left(q_{1}^{\ast}(j), q_{2}^{\ast}(j) \right), \forall j \in \mathcal{J} \right\rbrace$ is the optimal solution for time intervals $j \in \mathcal{J}$, then there exist $\lambda_{i}, i \in \{1, \cdots, 6j_{max} \}$, such that:
{\footnotesize
\[
(\widehat{\text{KKT}}) \left\{
\begin{array}{l}
\forall j \in \mathcal{J},\\
\quad 
\begin{array}{l}
\lambda_{i+6(j-1)} \geq 0, \quad \forall i \in \{1,2\ldots 6\}, \\
\\
\lambda_{1+6(j-1)} \cdot\left( \sum_{\tau=1}^{j}q_{1}^{\ast}(\tau)\Delta t_{\tau} - \bar{N}_{1}(j) \right) = 0,\\
\lambda_{2+6(j-1)} \cdot\left( \sum_{\tau=1}^{j}q_{2}^{\ast}(\tau)\Delta t_{\tau} - \bar{N}_{2}(j)\right) = 0, \\
\lambda_{3+6(j-1)} \cdot\left( \sum_{\tau=1}^{j}\left(q_{1}^{\ast}(\tau) + q_{2}^{\ast}(\tau) \right)\Delta t_{\tau} - \bar{N}_{3}(j) \right) = 0, \\
\lambda_{4+6(j-1)} \cdot \left( q_{1}^{\ast}(j) - q_{1}^{max}\right) = 0, \\
\lambda_{5+6(j-1)} \cdot \left( q_{2}^{\ast}(j) - q_{2}^{max} \right) = 0, \\
\lambda_{6+6(j-1)} \cdot \left( q_{1}^{\ast}(j) + q_{2}^{\ast}(j)  - q_{3}^{max} \right) = 0, \\
\\
-\frac{\partial f}{\partial q_{1}(j)\Delta t_{j}} + \sum_{l = j}^{j_{max}}(\lambda_{1+6(l-1)} + \lambda_{3+6(l-1)}) + \lambda_{4+6(j-1)} + \lambda_{6+6(j-1)} = 0, \\
-\frac{\partial f}{\partial q_{2}(j)\Delta t_{j}} + \sum_{l = j}^{j_{max}}(\lambda_{2+6(l-1)} + \lambda_{3+6(l-1)}) + \lambda_{5+6(j-1)} + \lambda_{6+6(j-1)} = 0. \\
\end{array}
\end{array}
\right.
\]
}

\emph{Remark:} There are six multipliers $\lambda_{i+6(j-1)}, i \in \{1,2,\dots, 6\}$ associated with each interval $j$. Except for the stationarity conditions, $\cup_{j\in \mathcal{J}} \text{KKT}_{j} $ is identical to $\widehat{\text{KKT}}$.\\

\textbf{\emph{Step 4: Show the optimal solution to CP~(\ref{CP:proofmultiplestep}) is the unique solution for all steps}}. 

Suppose $\left(q_{1}(j), q_{2}(j)\right), \forall j \in \mathcal{J} $ is the optimal solution to CP~(\ref{CP:proofmultiplestep}), then there exists a set of multipliers $\{\lambda_{i}\}$ that satisfies $\widehat{\text{KKT}}$. We now show that, the set of multipliers $\{\lambda_{i}\}$ also satisfies $\cup_{j=1}^{j_{max}} \text{KKT}_{j}$.

Since $\widehat{\text{KKT}}$ and $\cup_{j=1}^{j_{max}} \text{KKT}_{j}$ are identical except for the stationarity conditions, it suffices to prove that at each time interval $j$, the corresponding multipliers satisfies KKT$_{j}$ the following inequalities which is derived in Step 2.
{\footnotesize
\[
\left\lbrace
\begin{array}{lll}
\lambda_{1+6(j-1)} + \lambda_{3+6(j-1)} + \lambda_{4+6(j-1)} + \lambda_{6+6(j-1)} > 0, & & \langle 1 \rangle \\
\lambda_{2+6(j-1)}  + \lambda_{3+6(j-1)}  + \lambda_{5+6(j-1)}  + \lambda_{6+6(j-1)}  > 0, && \langle 2 \rangle\\
\lambda_{1+6(j-1)}  + \lambda_{3+6(j-1)}  + \lambda_{4+6(j-1)}  + \lambda_{6+6(j-1)}  > &\\   \qquad \qquad \lambda_{2+6(j-1)}  + \lambda_{3+6(j-1)}  + \lambda_{5+6(j-1)}  + \lambda_{6+6(j-1)}  & \text{ if } q_{2}(j) \geq Pq_{1}(j), & \langle 3 \rangle \\
\lambda_{1+6(j-1)}  + \lambda_{3+6(j-1)}  + \lambda_{4+6(j-1)}  + \lambda_{6+6(j-1)}  < & \\ \qquad \qquad \lambda_{2+6(j-1)}  + \lambda_{3+6(j-1)}  + \lambda_{5+6(j-1)}  + \lambda_{6+6(j-1)}  & \text{ if } q_{2}(j) < Pq_{1}(j). & \langle 4 \rangle\\
\end{array}
\right.
\]
}

\begin{itemize}
\item First show $\langle 1 \rangle$ is true.\\
When $j < j_{max}$,
{\footnotesize
\[
\begin{array}{r}
\frac{\partial f}{\partial q_{1}(j)\Delta t_{j}}  -\frac{\partial f}{\partial q_{1}(j+1)\Delta t_{j+1}} = \lambda_{1+6(j-1)} + \lambda_{3+6(j-1)} + \lambda_{4+6(j-1)} + \lambda_{6+6(j-1)}\\
 - \lambda_{4+6j} - \lambda_{6+6j}.
\end{array}
\]
}
Then, rearranging terms,
{\footnotesize
\[
\begin{array}{ll}
\lambda_{1+6(j-1)} + \lambda_{3+6(j-1)} + & \lambda_{4+6(j-1)} + \lambda_{6+6(j-1)} \\
& = \lambda_{4+6j} + \lambda_{6+6j} + \frac{\partial f}{\partial q_{1}(j)\Delta t_{j}}  -\frac{\partial f}{\partial q_{1}(j+1)\Delta t_{j+1}}, \\
& \geq \frac{\partial f}{\partial q_{1}(j)\Delta t_{j}}  -\frac{\partial f}{\partial q_{1}(j+1)\Delta t_{j+1}}, \\
& > 0, \qquad \text{by~\eqref{eq:prop_multiple_step_merge_1}}.
\end{array}
\]
}
When $j=j_{max}$, 
{\footnotesize
\[
\begin{array}{r}
\lambda_{1+6(j_{max}-1)} + \lambda_{3+6(j_{max}-1)} + \lambda_{4+6(j_{max} -1)} + \lambda_{6+6(j_{max} -1)}  = \frac{\partial f}{\partial q_{1}(j)\Delta t_{j}} > 0,\\
 \text{by~\eqref{eq:prop_multiple_step_merge_1}}.\\
 \end{array}
\]
}
\item $\langle 2 \rangle$ can be proved similarly as $\langle 1 \rangle$.\\
\item Next, we show $\langle 3 \rangle$ is true.\\
When $j < j_{max}$, and $q_{2}(j) \geq Pq_{1}(j)$,
{\footnotesize
\[
\begin{array}{ll}
\frac{\partial f}{\partial q_{1}(j)\Delta t_{j}}  -\frac{\partial f}{\partial q_{2}(j)\Delta t_{j}} &= \sum_{u=j+1}^{j_{max}}\left( \lambda_{1+6(u-1)} + \lambda_{3+6(u-1)} \right) + \lambda_{1+6(j-1)} + \lambda_{3+6(j-1)}\\
& + \lambda_{4+6(j-1)} + \lambda_{6+6(j-1)} - \sum_{u=j+1}^{j_{max}}\left( \lambda_{2+6(u-1)} + \lambda_{3+6(u-1)}\right) \\
&- \lambda_{2+6(j-1)} - \lambda_{3+6(j-1)} - \lambda_{5+6(j-1)} - \lambda_{6+6(j-1)}.
\end{array}
\]
}
Hence, 
{\footnotesize
\[
\begin{array}{l}
\left( \lambda_{1+6(j-1)} + \lambda_{3+6(j-1)} + \lambda_{4+6(j-1)} + \lambda_{6+6(j-1)} \right)\\ 
\qquad \qquad -\left( \lambda_{2+6(j-1)} + \lambda_{3+6(j-1)} + \lambda_{5+6(j-1)} + \lambda_{6+6(j-1)} \right)\\ 
= \frac{\partial f}{\partial q_{1}(j)\Delta t_{j}}  -\frac{\partial f}{\partial q_{2}(j)\Delta t_{j}} + \sum_{u=j+1}^{j_{max}}\left( \lambda_{2+6(u-1)} \right) -\sum_{u=j+1}^{j_{max}}\left( \lambda_{1+6(u-1)} \right), \\
 \geq \frac{\partial f}{\partial q_{1}(j)\Delta t_{j}}  -\frac{\partial f}{\partial q_{2}(j)\Delta t_{j}} -\sum_{u=j+1}^{j_{max}}\left( \lambda_{1+6(u-1)} \right), \\
 \geq \frac{\partial f}{\partial q_{1}(j)\Delta t_{j}}  -\frac{\partial f}{\partial q_{2}(j)\Delta t_{j}} - \frac{\partial f}{\partial q_{1}(j+1)\Delta t_{j+1}}, \\
 > 0, \quad \text{by~\eqref{eq:prop_multiple_step_merge_3}}.
\end{array}
\]
}
When $j = j_{max}$, and $q_{2}(j) \geq Pq_{1}(j)$,
\[
\begin{array}{l}
 \lambda_{1+6(j-1)} + \lambda_{3+6(j-1)} + \lambda_{4+6(j-1)} + \lambda_{6+6(j-1)} \\  
 \qquad \qquad - \lambda_{2+6(j-1)} - \lambda_{3+6(j-1)} - \lambda_{5+6(j-1)} - \lambda_{6+6(j-1)} \\
\qquad = \frac{\partial f}{\partial q_{1}(j_{max})\Delta t_{j_{max}}}  -\frac{\partial f}{\partial q_{2}(j_{max})\Delta t_{j_{max}}} >0, \qquad \text{by~\eqref{eq:prop_multiple_step_merge_5}}.
\end{array}
\]
\item The proof of $\langle 4 \rangle$ follows similarly as $\langle 3 \rangle$.\\
\end{itemize}

\end{proof}


\bibliographystyle{siamplain}
\bibliography{TrafficNetwork}

\end{document}